\documentclass[notitlepage,superscriptaddress,twocolumn,showpacs,nobalancelastpage,aps,prl]{revtex4-1}
\usepackage[english]{babel}
\RequirePackage[T1]{fontenc}
\RequirePackage{times} 

\usepackage{amsfonts}
\usepackage{subfigure}
\usepackage{amsmath}
\usepackage{amssymb}
\usepackage{bm}
\usepackage{graphicx,epstopdf}
\usepackage{appendix}
\usepackage{array}
\usepackage{color}
\usepackage[hidelinks]{hyperref}
\usepackage{cancel}
\usepackage{my_symbols}
\usepackage[normalem]{ulem}
\usepackage{CJK}
\providecommand{\U}[1]{\protect\rule{.1in}{.1in}}

{\par}

\newcommand\rmv{\bgroup\markoverwith {\textcolor{red}{\rule[0.5ex]{2pt}{0.4pt}}}\ULon}

\begin{document}
\begin{CJK*}{UTF8}{gbsn} 
\title{Torque-induced dispersive readout in a weakly coupled hybrid system}
\author{Vahram L. Grigoryan}
\affiliation{Institute for Quantum Science and Engineering, Southern University of Science and Technology, Shenzhen 518055, China}
\affiliation{Center for Quantum Computing, Peng Cheng Laboratory, Shenzhen 518005, China}
\author{Ke Xia}
\email[Corresponding author:~]{kexia@bnu.edu.cn}
\affiliation{Institute for Quantum Science and Engineering, Southern University of Science and Technology, Shenzhen 518055, China}
\affiliation{Center for Quantum Computing, Peng Cheng Laboratory, Shenzhen 518005, China}

\begin{abstract}
We propose a quantum state readout mechanism of a weakly coupled qubit in dispersive regime. The hybrid system consists of ferromagnetic insulator and a superconducting qubit in a microwave cavity. The enhancement of the measurement sensitivity is achieved by exerting torque on the ferromagnetic insulator magnetization, which compensates the damping of the system leading to an exceptional point. The proposed machanism allows to measure the qubit state either via the transmission of the cavity or the FMR signal of the magnetic material.
\end{abstract}
	\maketitle
\end{CJK*}

Quantum information processing is based on the storage, manipulation and readout of the state in quantum bit (qubit) \cite{nielsen_2000,bennett_2000}. In strongly coupled superconducting (SC) qubit and a cavity system, non-distructive readout of the quantum state can be realized in dispersive regime \cite{blais_2004,wallraff_2005,filipp_2009,vijay_2012}, where sufficient detuning of the cavity and the qubit frequencies prevents exchange of excitations between the systems. The readout of the quantum state of the qubit is based on qubit-state dependent frequency shift of the cavity due to virtual transitions, which can be measured from the transmission spectrum of the cavity. This measurement method is used not only for readout of SC qubit state \cite{wallraff_2004}. It has been applied to ac-driven quantum systems \cite{kohler_2017} and semiconductor quantum dots \cite{frey_2012,petersson_2012}. The shortcoming of the model, however, is that strong coupling (larger than the damping of the system) between the cavity and the SC qubit is required \cite{wallraff_2004} in order to see the frequency shift. This makes challenging to use the method for other systems such as a single electron spin, which are weakly coupled  to the cavity \cite{xiang_2013,kurizki_2015}.

From the other hand, realization of coherent \cite{bai_2015,huebl_2013,tabuchi_2014,zhang_2014,goryachev_2014} and dissipative magnon-photon \cite{grigoryan_2018,harder_2018,grigoryan_2019,bhoi_2019,boventer_2019,yu_2019} coupling in a microwave cavity, cavity-mediated dissipative magnon-magnon coupling \cite{grigoryan_2019_1,xu_2019}, magnon-photon entanglement \cite{yuan_2019} and other phenomena  push the boundaries of application of cavity spintroics. Moreover, recent demonstration \cite{li_2019,hou_2019} of strong coupling of cavity photons with on-chip nanomagnet magnetization opens new avenue for exploiting cavity-spintronics in quantum information technologies.

 Very recently, quantum state measurement in weakly coupled regime was proposed \cite{troiani_2019} using pair of near resonant modes in a cavity, where one of the mode is weakly coupled to the SC qubit and both cavity modes are coupled to the probe field. Another mechanism proposed to measure the quantum state is based on exploiting two coupled cavities with balanced gain and loss \cite{quijandria_2018,zhang_2019}. Tuning the gain from one of the cavities an exceptional point (EP) can be achieved in another cavity containing the SC qubit. Such EPs occur in non-Hermitian systems, where, by tuning some parameters of the system, a singularity point in eigenfunctions and eigenvalues can be revealed \cite{heiss_2012,grigoryan_2018,grigoryan_2019}.  Beside various exciting properties of such parity-time ($\mathcal{PT}$)-symmetric systems at EPs, such as the unidirectional invisibility \cite{feng_2013,lu_2015}, monochromatic microwave generation \cite{grigoryan_2019}, enhanced spontaneous emission \cite{lin_2016} and lasing \cite{liertzer_2012,feng_2014,hodaei_2014}, the sensitivity of the detection can also be enhanced near an EP in microcavity sensors \cite{wiersig_2014}.

\begin{figure}[t!]

		\includegraphics[width=\columnwidth]{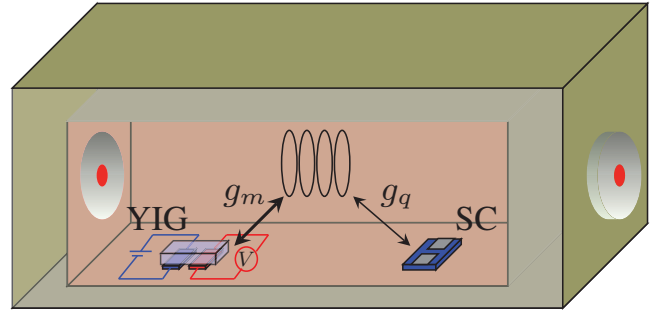}
\caption{The schematic picture of the system. YIG film and a SC qubit reside near the antinode of the magnetic field and electric field, respectively. $g_m$ and $g_q$ denote magnon-photon and qubit-photon couplings, respectively. The two ports on both sides of the cavity are for phtoton input and output. One of the strip lines (blue in the picture) patterned on the YIG film is used to exert torque on the YIG. Another strip line (red in the picture) is used to measure spin pumping voltage.}\label{fig:schem}
\end{figure}
Here we propose a mechanism of electric dispersive measurement of quantum state of a SC qubit in weak coupling regime. Our system consists of SC qubit and an yttrium iron garnet (YIG) film in a cavity. Here, the YIG magnetization is strongly coupled with the cavity photons, while the coupling of cavity with SC qubit is weak. The energy loss of the system is compensated via energy input into the magnon subsystem of YIG by applying torque. Varying the torque, coalescence or real and imaginary components of level energy is reached at EP when the gain exactly compensates the losses of the system \cite{grigoryan_2019}. We show that readout sensitivity is highly enhanced near the EP, which makes possible of electric and/or optical measurement of the SC quantum state. Exploiting the strong coupling between YIG and the cavity \cite{huebl_2013,tabuchi_2014,zhang_2014,goryachev_2014,bai_2015} it is possible to control and make measurements using YIG without disturbing the qubit state. Because the qubit is out of resonance with the cavity and not coupled with the YIG magnetization directly the measurement does not affect the qubit state. Moreover, exerting torque on the YIG allows us to use only one cavity mode and avoid complicated setups using coupled cavities \cite{zhang_2019} or different modes \cite{troiani_2019}.

So far, experiments in this field were performed by measuring the qubit state-dependent frequency shift of either the transmission ($S_{21}$) or reflection coefficient ($S_{11}$) of the microwave cavity \cite{blais_2004,wallraff_2005,filipp_2009,vijay_2012,schreier_2015}. We propose a technique, where spin pumping \cite{tserkovnyak_2005} enables electrical detection of quantum state in hybrid SC qubit-YIG system in a cavity. In \Figure{fig:schem} we show the schematic picture of our setup. The YIG film and the SC qubit are located in the cavity. The qubit weakly couples to the electric field of the cavity mode. The YIG film is mounted near the antinode of the magnetic field of the mode. A local magnetic field is applied on YIG, which makes the film a single-domain ferromagnet. Due to a large magnetization of the YIG film it couples strongly to the magnetic field of the cavity mode. Two platinum (Pt) layers are patterned on one side of the YIG film. One of these layers is used for exerting torque on the YIG film. The mechanisms to create "negative damping" or torque include spin Seebeck effect (SSE)induced spin transfer torque (STT) \cite{uchida_2010,holanda_2017,safranski_2017,grigoryan_2019} or spin Hall effect (SHE)-STT \cite{chen_2016,hamadeh_2014,sklenar_2015}. The second field is for spin pumping measurements \cite{tserkovnyak_2005} which enables electrical detection of the qubit state. The proposed setup enables both transmission measurements of the cavity and the electrical detection of FMR on samples loaded in the cavity.
\begin{figure}[t!]
\begin{tabular}{cc}
		\includegraphics[width=.49\columnwidth]{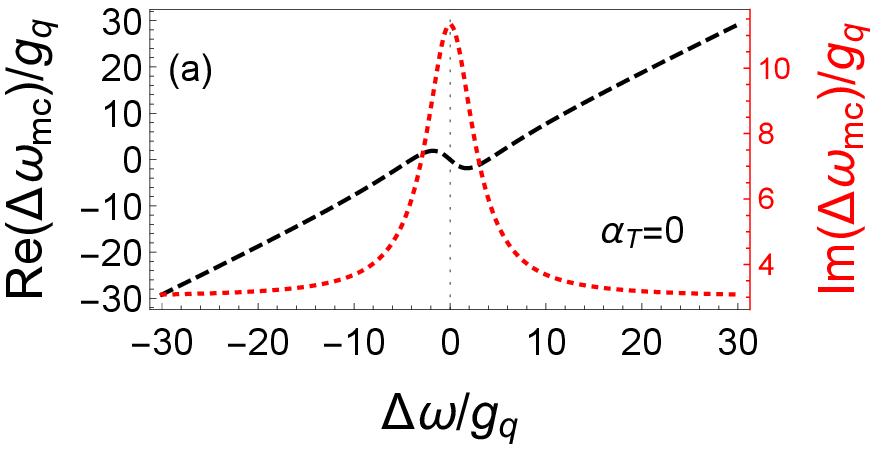}&\includegraphics[width=.49\columnwidth]{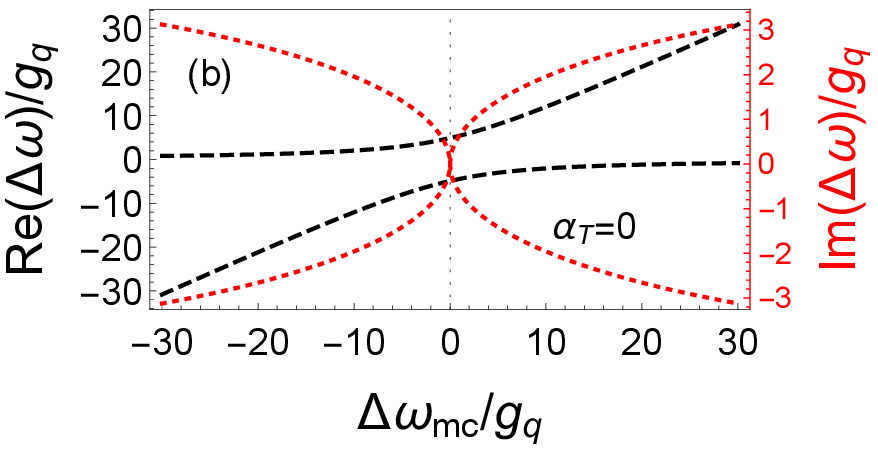}
		\\	\includegraphics[width=.49\columnwidth]{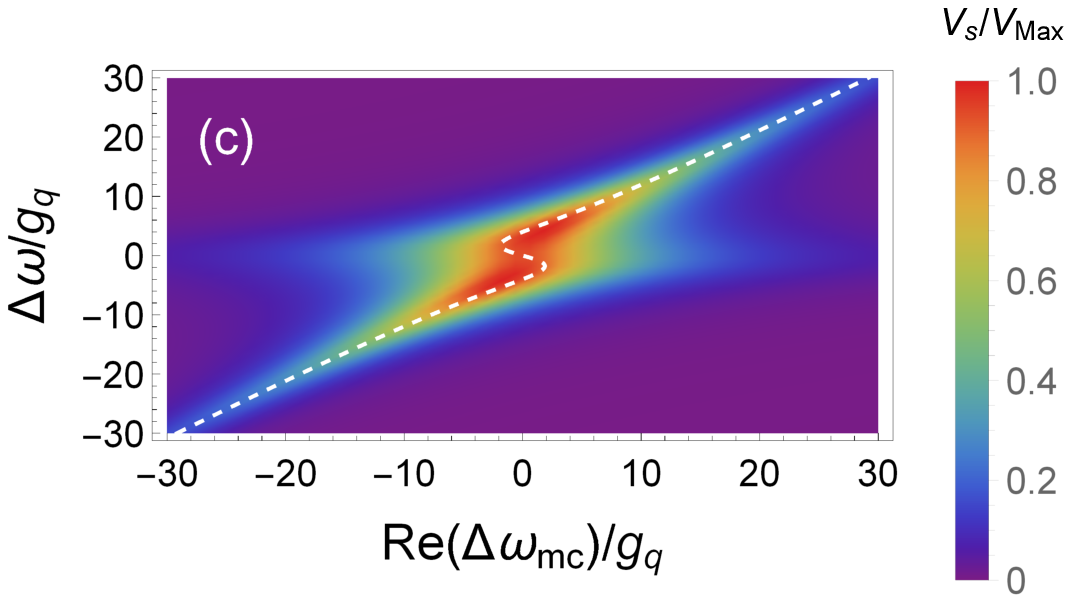}&
		\includegraphics[width=.49\columnwidth]{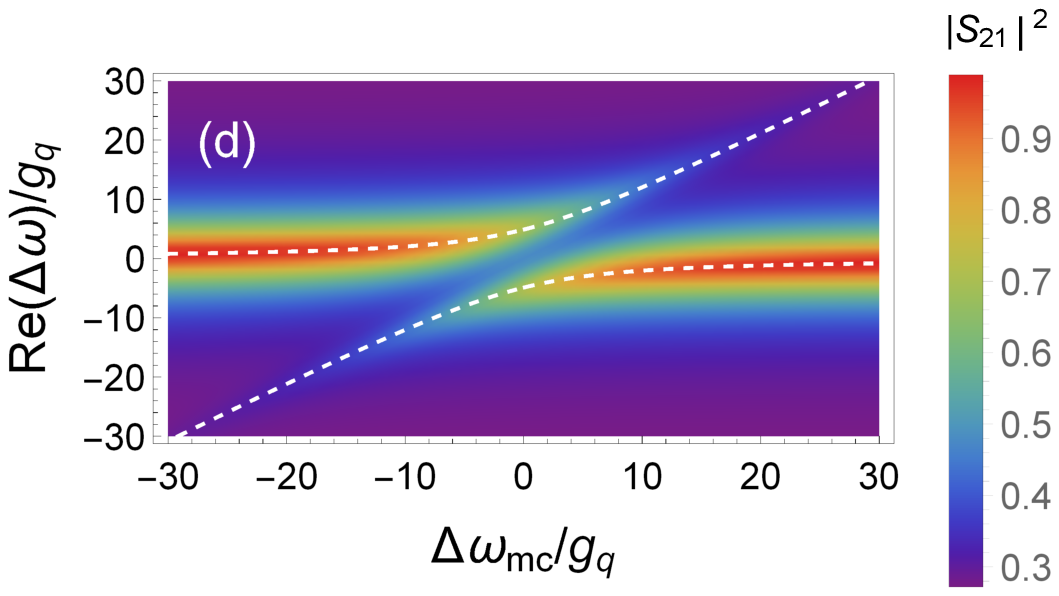}\\
	\includegraphics[width=.49\columnwidth]{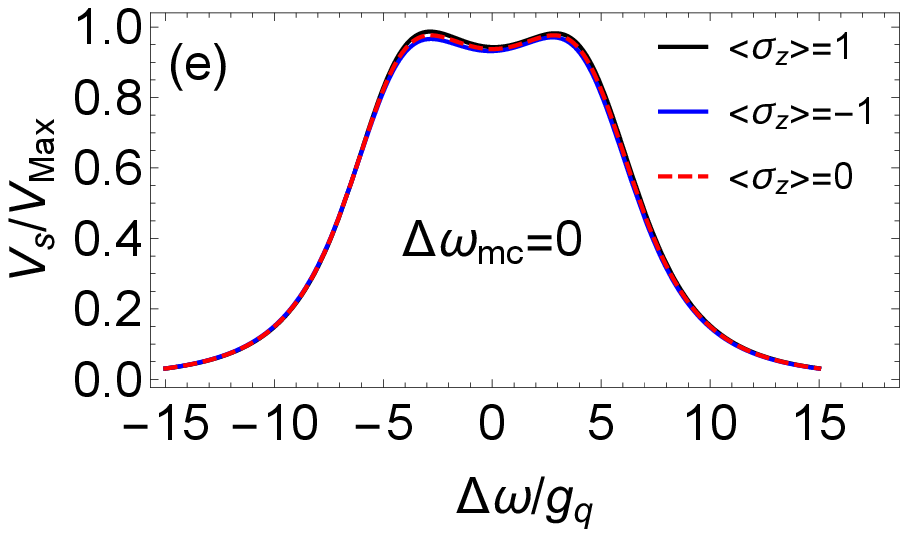}&
		\includegraphics[width=.49\columnwidth]{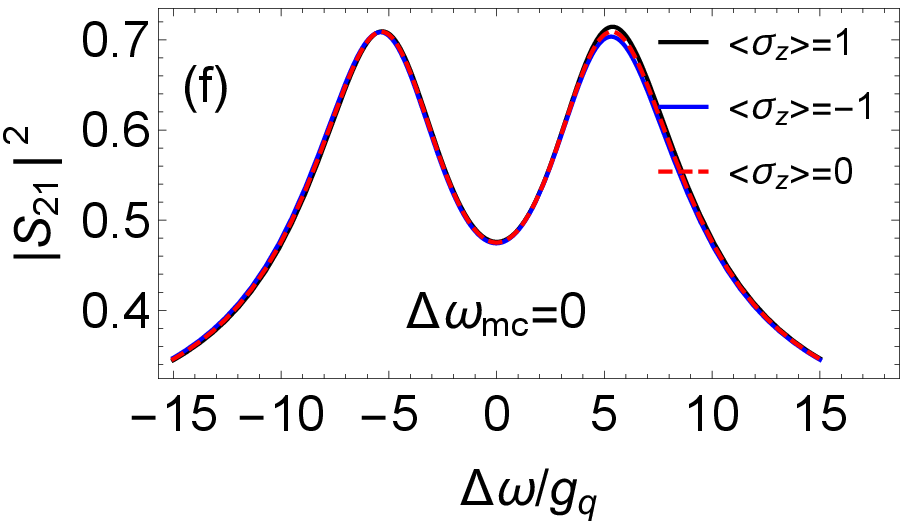}	
		\end{tabular}
\caption{(a) The left and right labels show the real (black dashed line) and imaginary (red dotted line) component of $\Delta \omega _{mc}$ as a function of $\Delta \omega ,$ respectively. (b) The same as in (a) for $\Delta\omega.$ (c) and (d) show the density plot of spin voltage and transmission amplitude, respectively. The dashed lines are the corresponding spectra. (e) and (f) depict the spin current and transmission amplitude evolution, respectively as a function of frequency $\Delta\omega$ at the value of $\Delta\omega_{mc}$ shown in the picture. The dashed red lines are in the absence of the qubit ($\left<\sigma_z\right>=0$), while the red and blue curves are for $\left<\sigma_z\right>=\pm 1.$ All pictures are in the absence of the torque on YIG magnetization ($\alpha _T=0$). The frequencies are presented in units of $g_q.$}\label{fig:2}
\end{figure}

\begin{figure}[h!]
\begin{tabular}{cc}
		\includegraphics[width=.49\columnwidth]{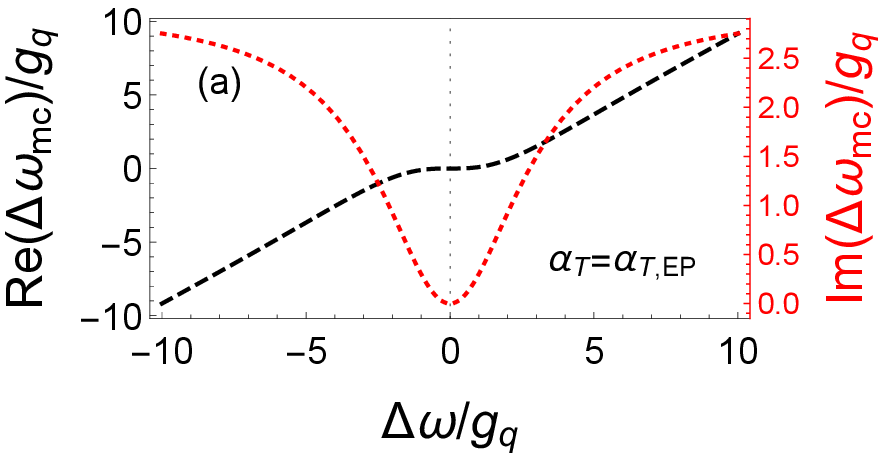}&
		\includegraphics[width=.49\columnwidth]{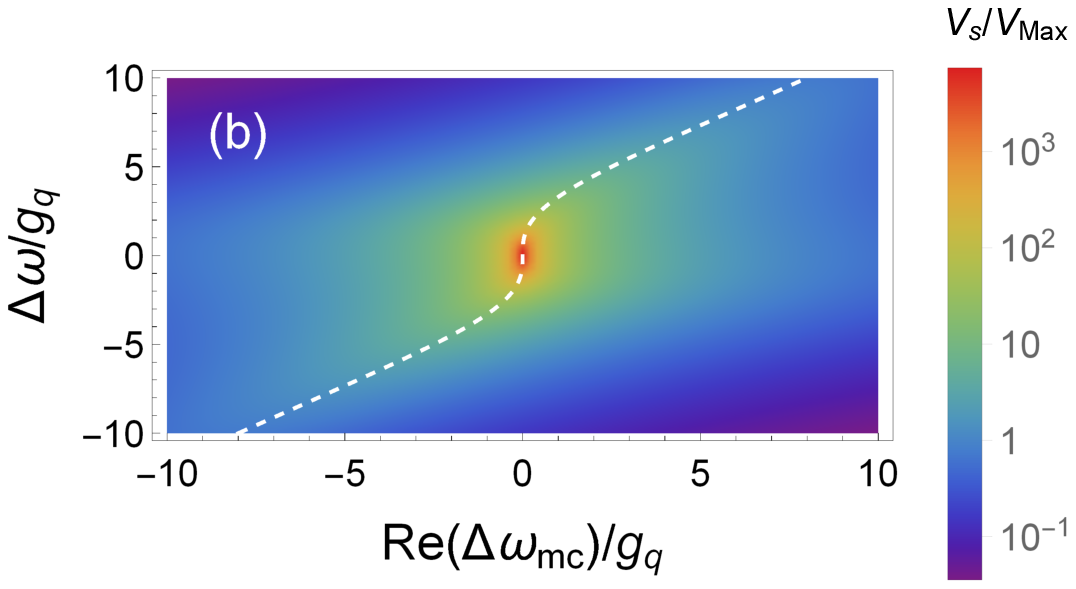}
	\\	\includegraphics[width=.49\columnwidth]{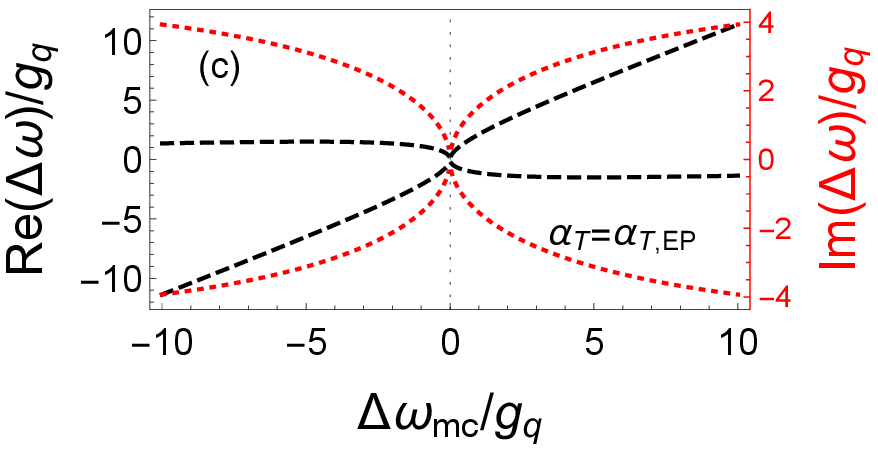}&
		\includegraphics[width=.49\columnwidth]{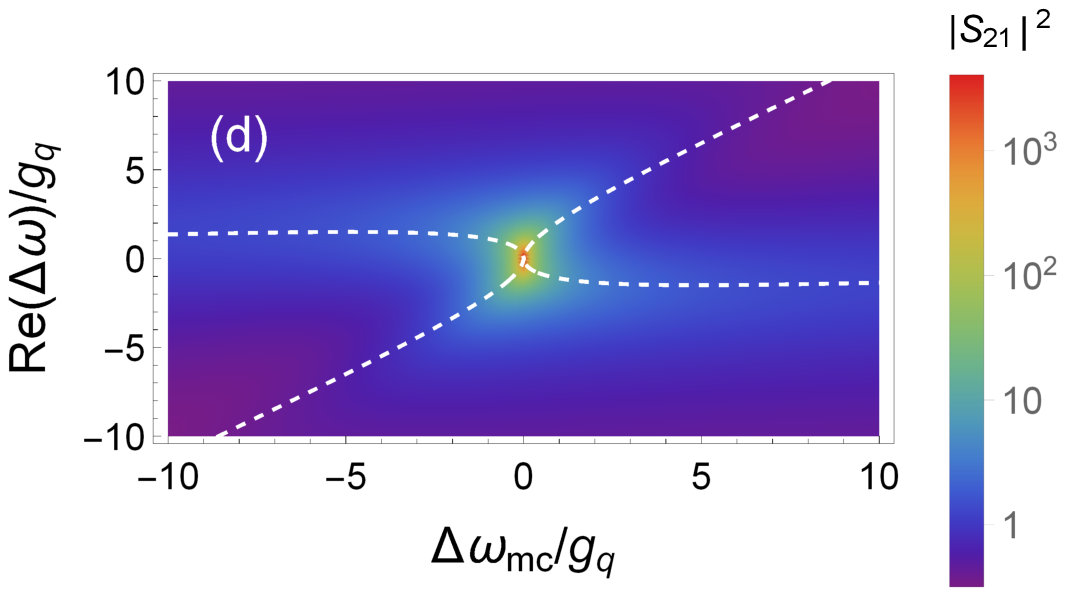}\\
		\includegraphics[width=.49\columnwidth]{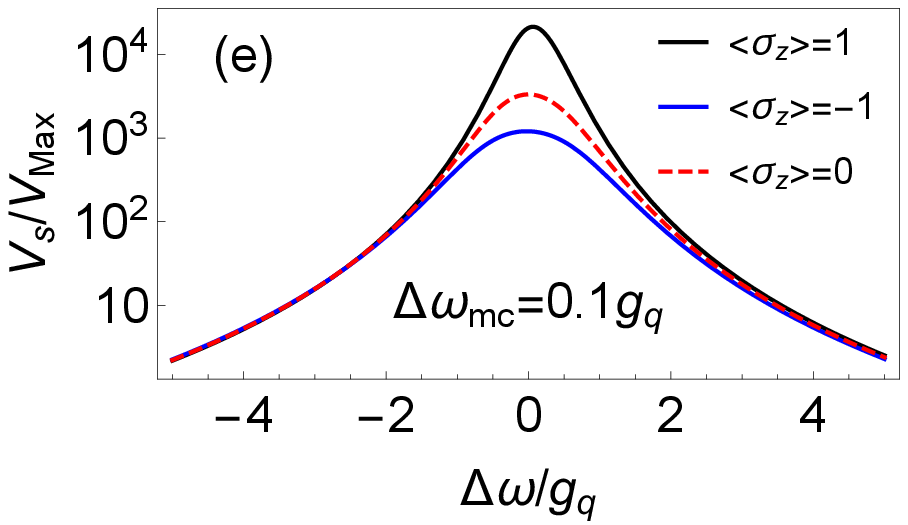}&\includegraphics[width=.49\columnwidth]{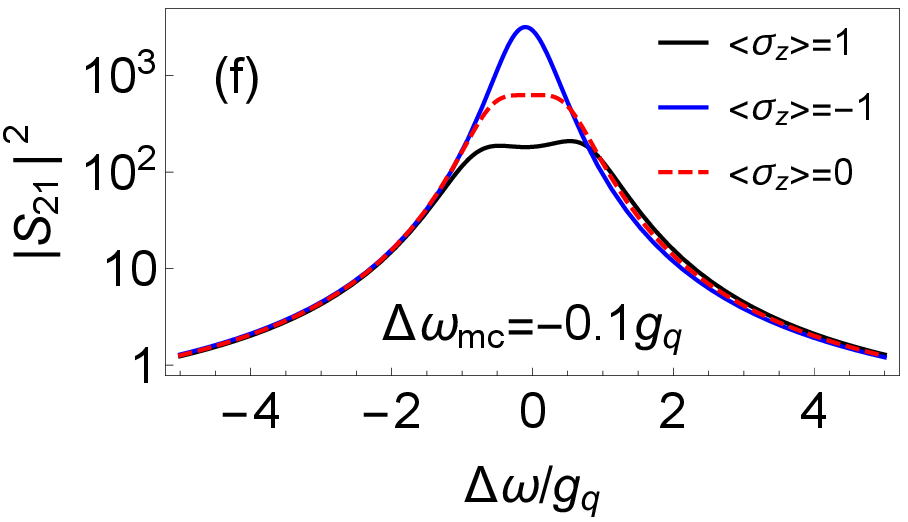}
		\end{tabular}
\caption{The same as in \Figure{fig:2} for the critical value of the spin torque $\alpha_T=\alpha_{T,EP}$.}\label{fig:3}
\end{figure}
\begin{figure*}[t!]
	\begin{tabular}{ccc}
			\includegraphics[width=.68\columnwidth]{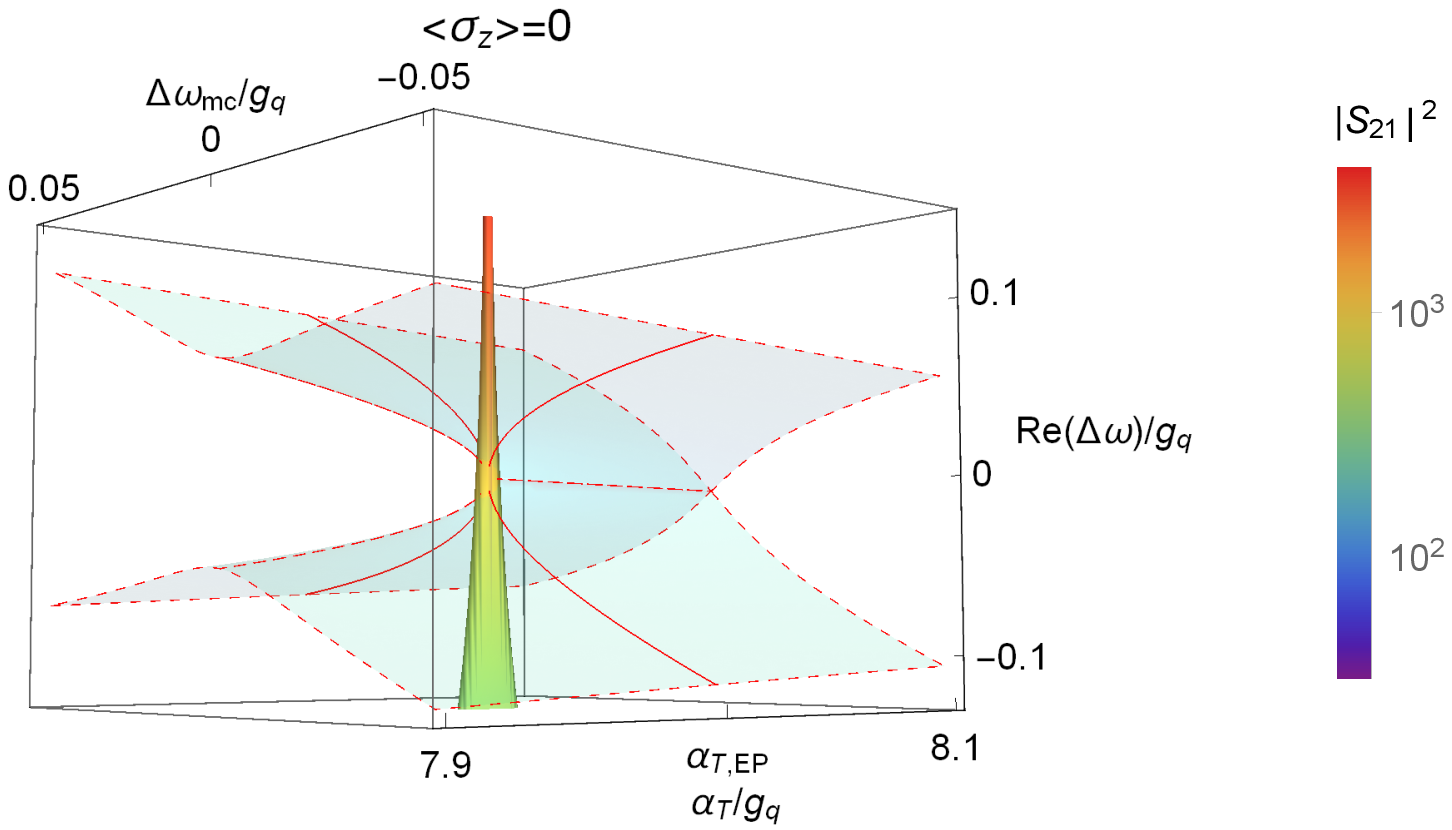}&
		\includegraphics[width=.62\columnwidth]{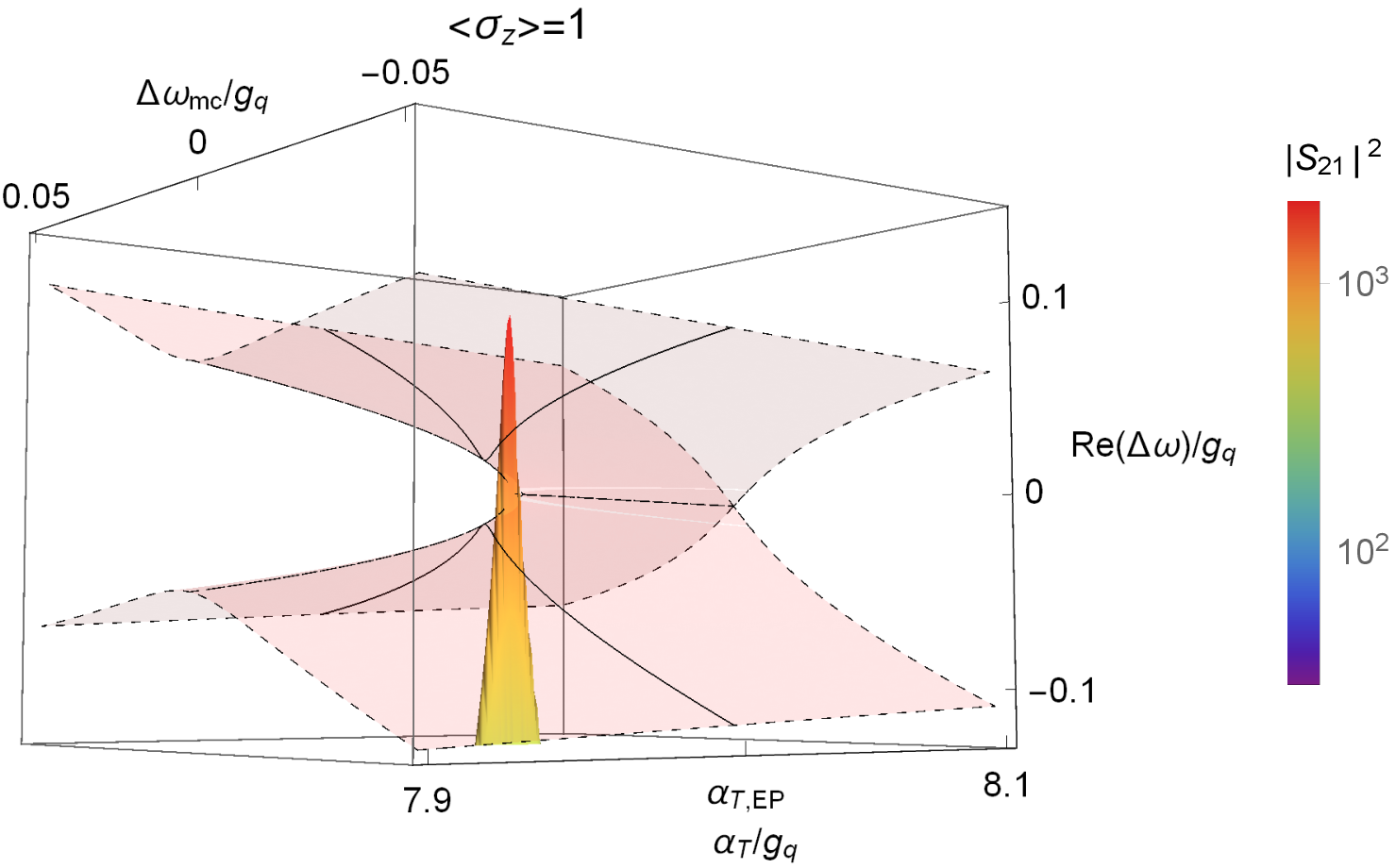}&
	\includegraphics[width=.62\columnwidth]{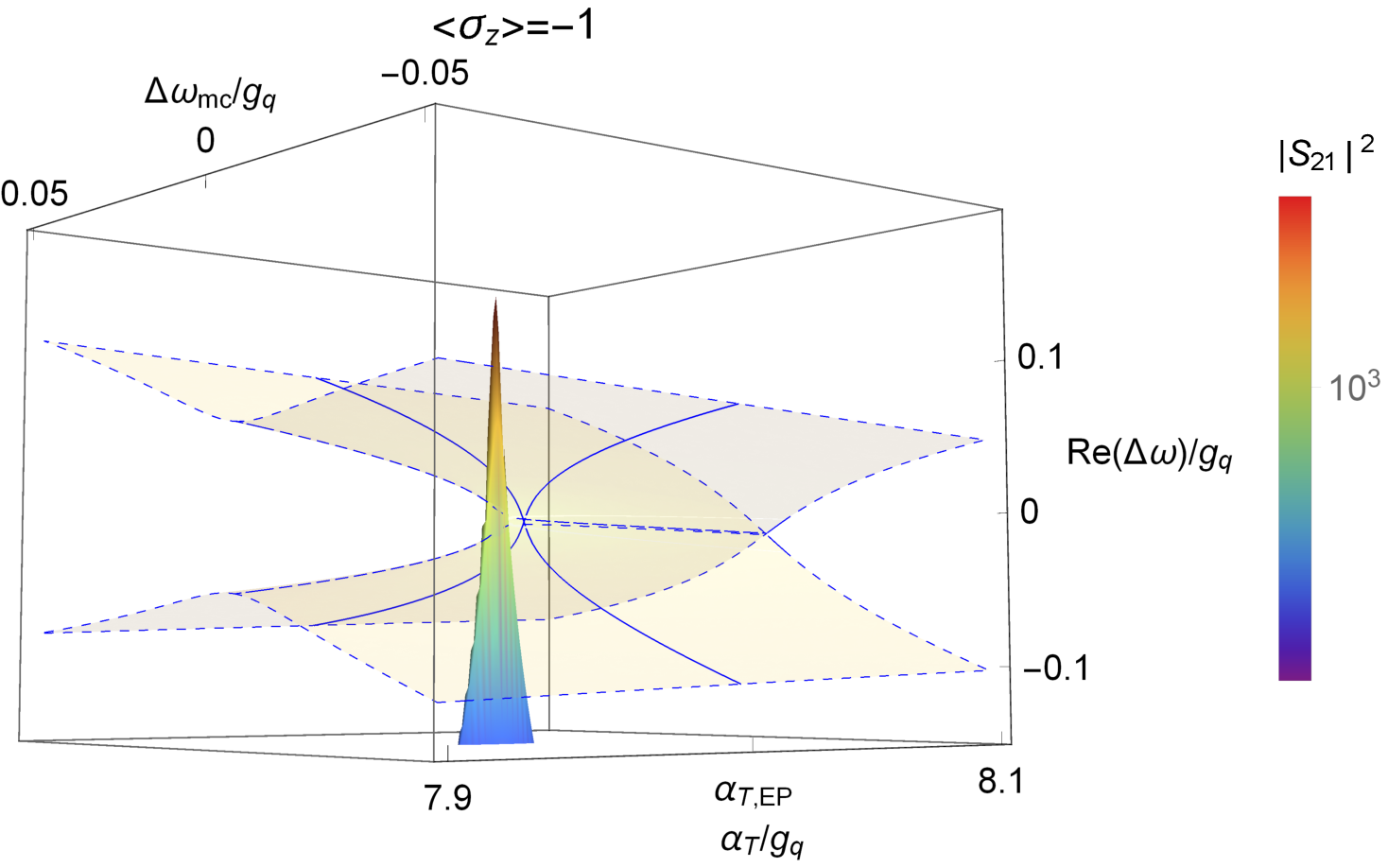}
	\end{tabular}
	\caption{The dependence of the real components of $\Delta\omega$ on $\Delta\omega_{mc}$ and the torque $\alpha_T$ for (a) $\left<\sigma_z\right>=0$, (b) $\left<\sigma_z\right>=1$ and (c) $\left<\sigma_z\right>=-1$. The lines show the spectrum for $\alpha_T=\alpha_{T,EP}.$ The colored cones show the transmission amplitude at $\Delta\omega_{mc}=\re{\Delta\omega}=0.$}\label{fig:4}
\end{figure*}
The system under study is governed by the following
Hamiltonian ($\hbar=1$)
\begin{align}
H_\ssf{sys}=&\omega_c a^\dag a+ {\omega_{q}\ov 2}\sigma_z+\omega_{m}m^\dag m\nn
&+ g_{q}\smlb{ \sigma^+ a+ a^\dag \sigma^-} +g_m \smlb{a^\dag m+a m^\dag} \label{eq:1},
\end{align}
where $a\smlb{a^\dag}$ and $m\smlb{m^\dag}$ are the cavity mode and magnon annihilation (creation) operators with frequency $\omega_c$ and $\omega_m,$ respectively. $\omega_q$ is the qubit transition frequency, $\sigma^\pm=\smlb{\sigma_x\pm i \sigma_y}/2$ are the ladder operators with $\sigma_x,$ $\sigma_y,$ and $\sigma_z$ being the spin-$1/2$ Pauli operators.
In rotating frame with respect to the cavity oscillation frequency $\omega_c$ the Hamiltonian \Eq{eq:1} becomes
\begin{align}
H_\ssf{sys}^\prime&=\Delta\omega_{mc} m^\dag m+{\Delta\omega_{qc}\ov 2}\sigma_z\nn
&+g_{q} \smlb{ \sigma^+ a + a^\dag\sigma^-} +g_m  \smlb{ a^\dag m  + a  m^\dag  } .\label{eq:2}
\end{align}
where $\Delta\omega_{mc}=\omega_m-\omega_c,~~\Delta\omega_{qc}=\omega_{q}-\omega_c.$ Based on \Eq{eq:2}, the quantum Langevin equations \cite{walls_1994} can be written as
\begin{align}
&\dot{a}=-ig_{q}  \sigma^- -ig_m    m-\kappa_a a +\sqrt{2\kappa_i}a_{in}\nn
&\dot{m}=-i\smlb{\Delta\omega_{mc}  m+g_m a}-\kappa_m m+\alpha_{T} m\nn
&\dot{\sigma}^-=-i\smlb{\Delta\omega_{qc}\sigma^- -g_{q} \sigma_z a  }-\gamma \sigma^- \label{eq:3},
\end{align}
where $\kappa_a,~\kappa_m,$ and $\gamma$ are the dissipations of cavity, magnetization and SC qubit, respectively. $\alpha_T$ is the torque on YIG spin \cite{grigoryan_2019}. $a_{in}$ is the input field of the cavity. Assuming the qubit to be in steady state ($\dot{\sigma}^-=0$) \cite{zhang_2019}
, we obtain from \Eq{eq:3}
\begin{align}
&  \sigma^- ={g_{q}  \ov \Delta\omega_{qc} -i\gamma} a\sigma_z. \label{eq:4}
\end{align}
After plugging \Eq{eq:4} back into \Eq{eq:3} we have
\begin{align}
&\dot{a}=-i\smlb{  {g_{q}^2  \ov \Delta\omega_{qc} -i\gamma}\sigma_z -i\kappa_a} a -ig_m    m +\sqrt{2\kappa_i}a_{in}\nn
&\dot{m}=-i\smlb{\Delta\omega_{mc}  m+g_m    a}-\kappa_m m+\alpha_T m \label{eq:5},
\end{align}
which can be used to recover the effective Hamiltonian of the system as
\begin{align}
H_{eff}&=\tilde{\omega}_c a^\dag a +\tilde{\omega}_m m^\dag m+ g_m\smlb{m^\dag a+a^\dag m}\label{eq:6}
\end{align}
where we made the following notations: $\tilde{\omega}_c \equiv {g_{q}^2  \ov \Delta\omega_{qc} -i\gamma}\sigma_z -i\kappa_a,$ and $\tilde{\omega}_m \equiv \Delta\omega_{mc}-i\smlb{\kappa_m-\alpha_T}.$ Diagonalizing the Hamiltonian \Eq{eq:6}, we obtain two eigenfrequencies in rotating frame $\Delta\omega\smlb{\Delta\omega_{mc}}$, where $\Delta\omega \equiv\omega-\omega_c$. The two positive real components of $\Delta\omega$ determine resonant frequencies, while imaginary parts describe damping of the coupled system. From the other hand, by solving the same equation, we obtain $\Delta\omega_{mc}$ as a function of $\Delta\omega$. In contrast to the solution for $\Delta\omega\smlb{\Delta\omega_{mc}}$, here we have only one solution. The real part of $\Delta\omega_{mc}\smlb{\Delta\omega}$ is the spectrum of ferromagnetic resonance (FMR) and the imaginary part is the linewidth \cite{grigoryan_2018,bai_2015}.

Next, we perform our calculations to derive spin pumping voltage as well as the transmission amplitude. From \Eq{eq:5} we obtain
\begin{widetext}
\begin{align}
&a=\frac{i\sqrt{2 \kappa _i}}{\left(\Delta \omega +i \kappa _{\alpha }\right)-\frac{g_m^2}{\Delta \omega +i \kappa _m-\Delta \omega _{\text{mc}}-i \alpha _T}-\frac{g_q^2 \sigma _z}{\Delta \omega _{\text{qc}}-i \gamma }} a_{\text{in}} \nn
&m=-\frac{i   g_m \sqrt{2\kappa _i}}{g_m^2+\left(\alpha _T+i \left(\Delta \omega +i \kappa _m-\Delta \omega _{\text{mc}}\right)\right) \left(-\kappa _{\alpha }+i \Delta \omega +\frac{g_q^2 \sigma _z}{\gamma +i \Delta \omega _{\text{qc}}}\right)}a_{\text{in}} \label{eq:7}
\end{align}
\end{widetext}
For spin pumping from YIG to attached metal layer we are interested in the ratio of magnetic excitation number ($N_m= m^\dag m$) and input photon number ($N_\text{in} = a^\dag _\text{in} a_\text{in}$) \cite{tserkovnyak_2005,walls_1994,clerk_2010}
\begin{equation}
V_\text{sp}\propto {N_m\ov N_\text{in}}={m^\dag m\ov a^\dag _\text{in} a_\text{in}} =\abs{m}^2\label{eq:8}.
\end{equation}
We also calculate the transmission using the input output theory, where we have \cite{walls_1994}
\begin{equation}
a_{in}=a_{out}+\sqrt{2\kappa_i}a \label{eq:9}.
\end{equation}
From \Eqs{eq:7}{eq:8} the transmission amplitude is
\begin{equation}
\abs{S_{21}}^2=\abs{a_\text{out}\ov a_\text{in}}^2.\label{eq:10}
\end{equation}

Our discussion starts with the results in the absence of torque on the magnetization ($\alpha_T = 0$). We use the following parameters for calculation: the damping rate of the cavity ($\kappa_\alpha$) and magnon $\kappa_m$ are $\kappa_\alpha=5 g_q,$ $\kappa_m=3g_q$ where $g_q$ is the coupling between cavity and superconducting qubit in weak coupling regime. The qubit decay rate is $\gamma=5 g_q.$ we set $\Delta\omega_{qc}=10g_q$, which ensures that the SC qubit is in dispersive regime. We set $\kappa_i=\kappa_m/2.$ In strong cavity-magnon coupling regime $g_m>\kappa_m,g_m=8g_q.$ In \Figure{fig:2} (a) we show the real and imaginary components of FMR spectrum ($\Delta\omega_{mc}$) as a function of $\Delta\omega$. The black dashed curve in the figure shows the FMR spectrum with varying slope at the region of coupling bandwidth. The red dotted curve shows the linewidth enhancement due to the magnon-photon coupling in the cavity \cite{bai_2015,grigoryan_2018}. Corresponding real and imaginary components of $\Delta\omega\smlb{\Delta\omega_{mc}}$ are shown in \Figure{fig:2} (b), where the usual level repulsion of the real components and coalescence of imaginary parts indicate magnon-photon coupling \cite{bai_2015,grigoryan_2018,huebl_2013,tabuchi_2014,zhang_2014,goryachev_2014}. The spin voltage (normalized by its maximum value) and the transmission amplitude dependence on frequencies $\Delta\omega$ and $\Delta\omega_{mc}$ are shown in Figs. \ref{fig:2} (c) and (d), respectively. The spin voltage and transmission in the figure are in-line with the experimental findings in YIG-cavity system \cite{bai_2015}. In Figs. \ref{fig:2} (e-f) we plot the spin voltage and transmission amplitude, respectively at $\Delta\omega_{mc}=0.$ The red dashed lines in the plots correspond to the case without SC qubit into system ($\left<\sigma_z\right>=0$). The black and blue lines show the spin pumping and transmission for $\left<\sigma_z\right>=\pm 1.$ It is seen from Figs. \ref{fig:2} (e-f) that it is hard to distinguish the SC qubit state from both, the spin volate and the transmission amplitude in weak coupling regime \cite{xiang_2013,kurizki_2015,zhang_2019}.

To measure the state of a weakly coupled qubit we now discuss the EP, where $\mathcal{PT}$-symmetry is breaking. We have shown previously \cite{grigoryan_2019} that the symmetry breaking and dramatic enhancement of the transmission and spin voltage occurs at resonance ($\Delta\omega_{mc}=0$) when the real components of $\Delta\omega$ coalesce and the damping of the system is compensated by the torque. This two conditions can be satisfied when $\alpha_{T}=\kappa_m+\kappa_\alpha\equiv \alpha_{T,EP}$ and $g_m=\kappa_\alpha$ \cite{grigoryan_2019,zhang_2019}. In \Figure{fig:3} we show the same plots as in \Figure{fig:2} near ($\Delta\omega=0.1g_q$) the $\mathcal{PT}$-symmetry breaking point, when $\alpha_T=\alpha_{T,EP}$ and $g_m=\kappa_\alpha.$ It is seen from \Figure{fig:3} (a) that the imaginary component of $\Delta\omega_{mc}$ becomes $0$ (red dotted curve) at the resonance condition ($\Delta\omega=0$), which indicates that the FMR damping is compensated \cite{grigoryan_2018,grigoryan_2019}. Moreover, as shown in \Figure{fig:3} (c), both imaginary components of $\Delta\omega$ become $0$ and coalesce at $\Delta\omega_{mc}=0.$ This feature, together with coalescence of the real components in \Figure{fig:3} (c) indicate the $\mathcal{PT}$-symmetry breaking at $\Delta\omega_{mc}=0.$ Fig. \ref{fig:3} (b) and (d) show the dramatic enhancement of the spin pumping and the transmission amplitude at the EP, which has been discussed in Ref. \onlinecite{grigoryan_2019}. In \Figure{fig:3} (e) we plot the dependence of the spin voltage near the EP (at $\Delta\omega_{mc}=0.1g_q$) as a function of $\Delta\omega$ for $\left<\sigma_z\right>=0$ (red dashed line) and $\left<\sigma_z\right>=\pm1$ (black and blue lines, respectively). It can be seen that the difference in spin voltage for different state of the SC qubit is prominant near the exceptional point. Similar results are shown in \Figure{fig:3} (f), where the transmission amplitude is plotted near the EP for different values of $\left<\sigma_z\right>.$ Similar to the spin voltage, the transmission amplitude is decreased by about one order of magnitude when the SC qubit is included with $\left<\sigma_z\right>=-1$ and increased when $\left<\sigma_z\right>=1.$

To understand the reason of difference in transmission amplitude for different qubit state, in \Figure{fig:4} (a-c) we plot the dependence of the eigenvalues of \Eq{eq:6} on $\Delta\omega_{mc}$ and the torque $\alpha_T.$ The transparent surface in the figures shows the spectrum for different value of $\alpha_T$, where the solid line shows the spectrum at EP ($\alpha_T=\alpha_{T,EP}$). It is seen that the system is $\mathcal{PT}$-symmetric for $\alpha_T< \alpha_{T,EP}$ and the symmetry is broken for $\alpha_T \geq \alpha_{T,EP}$.  The colored cones stand for the transmission amplitude at $\Delta\omega_{mc}=\re{\Delta\omega}=0$ for varying torque. For $\left<\sigma_z\right>=0$ in \Figure{fig:4} the transmission amplitude peak appears at $\alpha_{T,EP}.$ However, as follows from \Eq{eq:6} both, the real and imaginary components of the effective cavity frequency are altered by the SC qubit. The result is that magnitude of the torque necessary to reach the EP is different for different state of the qubit due to the change in the imaginary component. Moreover, the magnitude of the transmission amplitude is changed by the change of the real component of the effective cavity frequency. Particularly, the torque for $\left<\sigma_z\right>=1$ is larger than that of $\left<\sigma_z\right>=0$ (\Figure{fig:4} (b)) and smaller for $\left<\sigma_z\right>=-1$ (\Figure{fig:4} (c)). Consequently, the position of the peaks of the transmission amplitude are different for different qubit states.
\begin{figure}[t!]
\includegraphics[width=\columnwidth]{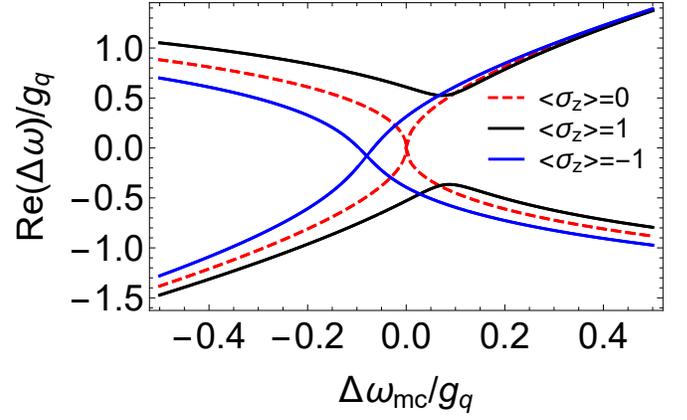}
\caption{The real component of the spectrum ($\Delta\omega(\Delta\omega_{mc})$ at the $\mathcal{PT}$-symmetry breaking point $\alpha_T=\alpha_{T,EP}$. The dashed, black and blue lines correspond to $\left<\sigma_z\right>=0,1,-1,$ respectively.}\label{fig:5}
\end{figure}

In \Figure{fig:5} we show the spectrum at $\alpha_T=\alpha_{T,EP}$ for different values of $\left<\sigma_z\right>.$ It follows that for the critical value of the torque $\alpha_{T,EP}$ the system is at the EP at $\Delta\omega_{mc}=0.$ When the qubit is introduced, the $\mathcal{PT}$-symmetry of the system is broken for $\left<\sigma_z\right>=-1$ while it is conserved when $\left<\sigma_z\right>=1$. Thus, depending on the qubit state the system can be switched form $\mathcal{PT}$-symmetric to broken $\mathcal{PT}$-symmetry states. 

In summary, we propose a mechanism of enhancing the sensitivity of SC qubit state measurement exploiting properties of $\mathcal{PT}$-symmetric systems. The dispersive readout mechanism in YIG-SC qubit-cavity hybrid system is based on compensating the damping of the systems using torque on the YIG magnetization. We demonstrate the enhancement of the sensitivity of readout near EP. Strong coupling between YIG and the cavity together with weak coupling of the qubit allows to control and make measurements using YIG without disturbing the qubit state in dispersive regime. Moreover, exerting torque on the YIG allows us to use only one cavity mode and avoid complicated setups using coupled cavities \cite{zhang_2019} or different modes \cite{troiani_2019}. The readout mechanism includes both transmission measurement of the cavity and electric measurement of the FMR signal in YIG. Beside the readout method the proposed mechanism allows realization of switching between $\mathcal{PT}$-symmetric to $\mathcal{PT}$-symmetry broken states via controlling the qubit state.


\begin{thebibliography}{49}%
\makeatletter
\providecommand \@ifxundefined [1]{%
 \@ifx{#1\undefined}
}%
\providecommand \@ifnum [1]{%
 \ifnum #1\expandafter \@firstoftwo
 \else \expandafter \@secondoftwo
 \fi
}%
\providecommand \@ifx [1]{%
 \ifx #1\expandafter \@firstoftwo
 \else \expandafter \@secondoftwo
 \fi
}%
\providecommand \natexlab [1]{#1}%
\providecommand \enquote  [1]{``#1''}%
\providecommand \bibnamefont  [1]{#1}%
\providecommand \bibfnamefont [1]{#1}%
\providecommand \citenamefont [1]{#1}%
\providecommand \href@noop [0]{\@secondoftwo}%
\providecommand \href [0]{\begingroup \@sanitize@url \@href}%
\providecommand \@href[1]{\@@startlink{#1}\@@href}%
\providecommand \@@href[1]{\endgroup#1\@@endlink}%
\providecommand \@sanitize@url [0]{\catcode `\\12\catcode `\$12\catcode
  `\&12\catcode `\#12\catcode `\^12\catcode `\_12\catcode `\%12\relax}%
\providecommand \@@startlink[1]{}%
\providecommand \@@endlink[0]{}%
\providecommand \url  [0]{\begingroup\@sanitize@url \@url }%
\providecommand \@url [1]{\endgroup\@href {#1}{\urlprefix }}%
\providecommand \urlprefix  [0]{URL }%
\providecommand \Eprint [0]{\href }%
\providecommand \doibase [0]{http://dx.doi.org/}%
\providecommand \selectlanguage [0]{\@gobble}%
\providecommand \bibinfo  [0]{\@secondoftwo}%
\providecommand \bibfield  [0]{\@secondoftwo}%
\providecommand \translation [1]{[#1]}%
\providecommand \BibitemOpen [0]{}%
\providecommand \bibitemStop [0]{}%
\providecommand \bibitemNoStop [0]{.\EOS\space}%
\providecommand \EOS [0]{\spacefactor3000\relax}%
\providecommand \BibitemShut  [1]{\csname bibitem#1\endcsname}%
\let\auto@bib@innerbib\@empty
\bibitem [{\citenamefont {Nielsen}(2000)}]{nielsen_2000}%
  \BibitemOpen
  \bibfield  {author} {\bibinfo {author} {\bibfnamefont {M.~A.}\ \bibnamefont
  {Nielsen}},\ }\href@noop {} {\emph {\bibinfo {title} {Quantum computation and
  quantum information}}}\ (\bibinfo  {publisher} {Cambridge University Press},\
  \bibinfo {address} {Cambridge ; New York},\ \bibinfo {year}
  {2000})\BibitemShut {NoStop}%
\bibitem [{\citenamefont {Bennett}\ and\ \citenamefont
  {Divincenzo}(2000)}]{bennett_2000}%
  \BibitemOpen
  \bibfield  {author} {\bibinfo {author} {\bibfnamefont {C.~H.}\ \bibnamefont
  {Bennett}}\ and\ \bibinfo {author} {\bibfnamefont {D.~P.}\ \bibnamefont
  {Divincenzo}},\ }\href@noop {} {\bibfield  {journal} {\bibinfo  {journal}
  {Nature}\ }\textbf {\bibinfo {volume} {404}},\ \bibinfo {pages} {247}
  (\bibinfo {year} {2000})}\BibitemShut {NoStop}%
\bibitem [{\citenamefont {Blais}\ \emph {et~al.}(2004)\citenamefont {Blais},
  \citenamefont {Huang}, \citenamefont {Wallraff}, \citenamefont {Girvin},\
  and\ \citenamefont {Schoelkopf}}]{blais_2004}%
  \BibitemOpen
  \bibfield  {author} {\bibinfo {author} {\bibfnamefont {A.}~\bibnamefont
  {Blais}}, \bibinfo {author} {\bibfnamefont {R.-S.}\ \bibnamefont {Huang}},
  \bibinfo {author} {\bibfnamefont {A.}~\bibnamefont {Wallraff}}, \bibinfo
  {author} {\bibfnamefont {S.~M.}\ \bibnamefont {Girvin}}, \ and\ \bibinfo
  {author} {\bibfnamefont {R.~J.}\ \bibnamefont {Schoelkopf}},\ }\href
  {\doibase 10.1103/PhysRevA.69.062320} {\bibfield  {journal} {\bibinfo
  {journal} {Phys. Rev. A}\ }\textbf {\bibinfo {volume} {69}},\ \bibinfo
  {pages} {062320} (\bibinfo {year} {2004})}\BibitemShut {NoStop}%
\bibitem [{\citenamefont {Wallraff}\ \emph {et~al.}(2005)\citenamefont
  {Wallraff}, \citenamefont {Schuster}, \citenamefont {Blais}, \citenamefont
  {Frunzio}, \citenamefont {Majer}, \citenamefont {Devoret}, \citenamefont
  {Girvin},\ and\ \citenamefont {Schoelkopf}}]{wallraff_2005}%
  \BibitemOpen
  \bibfield  {author} {\bibinfo {author} {\bibfnamefont {A.}~\bibnamefont
  {Wallraff}}, \bibinfo {author} {\bibfnamefont {D.~I.}\ \bibnamefont
  {Schuster}}, \bibinfo {author} {\bibfnamefont {A.}~\bibnamefont {Blais}},
  \bibinfo {author} {\bibfnamefont {L.}~\bibnamefont {Frunzio}}, \bibinfo
  {author} {\bibfnamefont {J.}~\bibnamefont {Majer}}, \bibinfo {author}
  {\bibfnamefont {M.~H.}\ \bibnamefont {Devoret}}, \bibinfo {author}
  {\bibfnamefont {S.~M.}\ \bibnamefont {Girvin}}, \ and\ \bibinfo {author}
  {\bibfnamefont {R.~J.}\ \bibnamefont {Schoelkopf}},\ }\href {\doibase
  10.1103/PhysRevLett.95.060501} {\bibfield  {journal} {\bibinfo  {journal}
  {Phys. Rev. Lett.}\ }\textbf {\bibinfo {volume} {95}},\ \bibinfo {pages}
  {060501} (\bibinfo {year} {2005})}\BibitemShut {NoStop}%
\bibitem [{\citenamefont {Filipp}\ \emph {et~al.}(2009)\citenamefont {Filipp},
  \citenamefont {Maurer}, \citenamefont {Leek}, \citenamefont {Baur},
  \citenamefont {Bianchetti}, \citenamefont {Fink}, \citenamefont {G\"oppl},
  \citenamefont {Steffen}, \citenamefont {Gambetta}, \citenamefont {Blais},\
  and\ \citenamefont {Wallraff}}]{filipp_2009}%
  \BibitemOpen
  \bibfield  {author} {\bibinfo {author} {\bibfnamefont {S.}~\bibnamefont
  {Filipp}}, \bibinfo {author} {\bibfnamefont {P.}~\bibnamefont {Maurer}},
  \bibinfo {author} {\bibfnamefont {P.~J.}\ \bibnamefont {Leek}}, \bibinfo
  {author} {\bibfnamefont {M.}~\bibnamefont {Baur}}, \bibinfo {author}
  {\bibfnamefont {R.}~\bibnamefont {Bianchetti}}, \bibinfo {author}
  {\bibfnamefont {J.~M.}\ \bibnamefont {Fink}}, \bibinfo {author}
  {\bibfnamefont {M.}~\bibnamefont {G\"oppl}}, \bibinfo {author} {\bibfnamefont
  {L.}~\bibnamefont {Steffen}}, \bibinfo {author} {\bibfnamefont {J.~M.}\
  \bibnamefont {Gambetta}}, \bibinfo {author} {\bibfnamefont {A.}~\bibnamefont
  {Blais}}, \ and\ \bibinfo {author} {\bibfnamefont {A.}~\bibnamefont
  {Wallraff}},\ }\href {\doibase 10.1103/PhysRevLett.102.200402} {\bibfield
  {journal} {\bibinfo  {journal} {Phys. Rev. Lett.}\ }\textbf {\bibinfo
  {volume} {102}},\ \bibinfo {pages} {200402} (\bibinfo {year}
  {2009})}\BibitemShut {NoStop}%
\bibitem [{\citenamefont {Vijay}\ \emph {et~al.}(2012)\citenamefont {Vijay},
  \citenamefont {Macklin}, \citenamefont {Slichter}, \citenamefont {Weber},
  \citenamefont {Murch}, \citenamefont {Naik}, \citenamefont {Korotkov},\ and\
  \citenamefont {Siddiqi}}]{vijay_2012}%
  \BibitemOpen
  \bibfield  {author} {\bibinfo {author} {\bibfnamefont {R.}~\bibnamefont
  {Vijay}}, \bibinfo {author} {\bibfnamefont {C.}~\bibnamefont {Macklin}},
  \bibinfo {author} {\bibfnamefont {D.~H.}\ \bibnamefont {Slichter}}, \bibinfo
  {author} {\bibfnamefont {S.~J.}\ \bibnamefont {Weber}}, \bibinfo {author}
  {\bibfnamefont {K.~W.}\ \bibnamefont {Murch}}, \bibinfo {author}
  {\bibfnamefont {R.}~\bibnamefont {Naik}}, \bibinfo {author} {\bibfnamefont
  {A.~N.}\ \bibnamefont {Korotkov}}, \ and\ \bibinfo {author} {\bibfnamefont
  {I.}~\bibnamefont {Siddiqi}},\ }\href@noop {} {\bibfield  {journal} {\bibinfo
   {journal} {Nature}\ }\textbf {\bibinfo {volume} {490}},\ \bibinfo {pages}
  {77} (\bibinfo {year} {2012})}\BibitemShut {NoStop}%
\bibitem [{\citenamefont {Wallraff}\ \emph {et~al.}(2004)\citenamefont
  {Wallraff}, \citenamefont {Schuster}, \citenamefont {Blais}, \citenamefont
  {Frunzio}, \citenamefont {Huang}, \citenamefont {Majer}, \citenamefont
  {Kumar}, \citenamefont {Girvin},\ and\ \citenamefont
  {Schoelkopf}}]{wallraff_2004}%
  \BibitemOpen
  \bibfield  {author} {\bibinfo {author} {\bibfnamefont {A.}~\bibnamefont
  {Wallraff}}, \bibinfo {author} {\bibfnamefont {D.~I.}\ \bibnamefont
  {Schuster}}, \bibinfo {author} {\bibfnamefont {A.}~\bibnamefont {Blais}},
  \bibinfo {author} {\bibfnamefont {L.}~\bibnamefont {Frunzio}}, \bibinfo
  {author} {\bibfnamefont {R.-S.}\ \bibnamefont {Huang}}, \bibinfo {author}
  {\bibfnamefont {J.}~\bibnamefont {Majer}}, \bibinfo {author} {\bibfnamefont
  {S.}~\bibnamefont {Kumar}}, \bibinfo {author} {\bibfnamefont {S.~M.}\
  \bibnamefont {Girvin}}, \ and\ \bibinfo {author} {\bibfnamefont {R.~J.}\
  \bibnamefont {Schoelkopf}},\ }\href@noop {} {\bibfield  {journal} {\bibinfo
  {journal} {Nature}\ }\textbf {\bibinfo {volume} {431}} (\bibinfo {year}
  {2004})}\BibitemShut {NoStop}%
\bibitem [{\citenamefont {Kohler}(2017)}]{kohler_2017}%
  \BibitemOpen
  \bibfield  {author} {\bibinfo {author} {\bibfnamefont {S.}~\bibnamefont
  {Kohler}},\ }\href {\doibase 10.1103/PhysRevLett.119.196802} {\bibfield
  {journal} {\bibinfo  {journal} {Phys. Rev. Lett.}\ }\textbf {\bibinfo
  {volume} {119}},\ \bibinfo {pages} {196802} (\bibinfo {year}
  {2017})}\BibitemShut {NoStop}%
\bibitem [{\citenamefont {Frey}\ \emph {et~al.}(2012)\citenamefont {Frey},
  \citenamefont {Leek}, \citenamefont {Beck}, \citenamefont {Blais},
  \citenamefont {Ihn}, \citenamefont {Ensslin},\ and\ \citenamefont
  {Wallraff}}]{frey_2012}%
  \BibitemOpen
  \bibfield  {author} {\bibinfo {author} {\bibfnamefont {T.}~\bibnamefont
  {Frey}}, \bibinfo {author} {\bibfnamefont {P.~J.}\ \bibnamefont {Leek}},
  \bibinfo {author} {\bibfnamefont {M.}~\bibnamefont {Beck}}, \bibinfo {author}
  {\bibfnamefont {A.}~\bibnamefont {Blais}}, \bibinfo {author} {\bibfnamefont
  {T.}~\bibnamefont {Ihn}}, \bibinfo {author} {\bibfnamefont {K.}~\bibnamefont
  {Ensslin}}, \ and\ \bibinfo {author} {\bibfnamefont {A.}~\bibnamefont
  {Wallraff}},\ }\href {\doibase 10.1103/PhysRevLett.108.046807} {\bibfield
  {journal} {\bibinfo  {journal} {Phys. Rev. Lett.}\ }\textbf {\bibinfo
  {volume} {108}},\ \bibinfo {pages} {046807} (\bibinfo {year}
  {2012})}\BibitemShut {NoStop}%
\bibitem [{\citenamefont {Petersson}\ \emph {et~al.}(2012)\citenamefont
  {Petersson}, \citenamefont {Mcfaul}, \citenamefont {Schroer}, \citenamefont
  {Jung}, \citenamefont {Taylor}, \citenamefont {Houck},\ and\ \citenamefont
  {Petta}}]{petersson_2012}%
  \BibitemOpen
  \bibfield  {author} {\bibinfo {author} {\bibfnamefont {K.~D.}\ \bibnamefont
  {Petersson}}, \bibinfo {author} {\bibfnamefont {L.~W.}\ \bibnamefont
  {Mcfaul}}, \bibinfo {author} {\bibfnamefont {M.~D.}\ \bibnamefont {Schroer}},
  \bibinfo {author} {\bibfnamefont {M.}~\bibnamefont {Jung}}, \bibinfo {author}
  {\bibfnamefont {J.~M.}\ \bibnamefont {Taylor}}, \bibinfo {author}
  {\bibfnamefont {A.~A.}\ \bibnamefont {Houck}}, \ and\ \bibinfo {author}
  {\bibfnamefont {J.~R.}\ \bibnamefont {Petta}},\ }\href@noop {} {\bibfield
  {journal} {\bibinfo  {journal} {Nature}\ }\textbf {\bibinfo {volume} {490}},\
  \bibinfo {pages} {380} (\bibinfo {year} {2012})}\BibitemShut {NoStop}%
\bibitem [{\citenamefont {Xiang}\ \emph {et~al.}(2013)\citenamefont {Xiang},
  \citenamefont {Ashhab}, \citenamefont {You},\ and\ \citenamefont
  {Nori}}]{xiang_2013}%
  \BibitemOpen
  \bibfield  {author} {\bibinfo {author} {\bibfnamefont {Z.-L.}\ \bibnamefont
  {Xiang}}, \bibinfo {author} {\bibfnamefont {S.}~\bibnamefont {Ashhab}},
  \bibinfo {author} {\bibfnamefont {J.~Q.}\ \bibnamefont {You}}, \ and\
  \bibinfo {author} {\bibfnamefont {F.}~\bibnamefont {Nori}},\ }\href {\doibase
  10.1103/RevModPhys.85.623} {\bibfield  {journal} {\bibinfo  {journal} {Rev.
  Mod. Phys.}\ }\textbf {\bibinfo {volume} {85}},\ \bibinfo {pages} {623}
  (\bibinfo {year} {2013})}\BibitemShut {NoStop}%
\bibitem [{\citenamefont {Kurizki}\ \emph {et~al.}(2015)\citenamefont
  {Kurizki}, \citenamefont {Bertet}, \citenamefont {Kubo}, \citenamefont
  {Molmer}, \citenamefont {Petrosyan}, \citenamefont {Rabl},\ and\
  \citenamefont {Schmiedmayer}}]{kurizki_2015}%
  \BibitemOpen
  \bibfield  {author} {\bibinfo {author} {\bibfnamefont {G.}~\bibnamefont
  {Kurizki}}, \bibinfo {author} {\bibfnamefont {P.}~\bibnamefont {Bertet}},
  \bibinfo {author} {\bibfnamefont {Y.}~\bibnamefont {Kubo}}, \bibinfo {author}
  {\bibfnamefont {K.}~\bibnamefont {Molmer}}, \bibinfo {author} {\bibfnamefont
  {D.}~\bibnamefont {Petrosyan}}, \bibinfo {author} {\bibfnamefont
  {P.}~\bibnamefont {Rabl}}, \ and\ \bibinfo {author} {\bibfnamefont
  {J.}~\bibnamefont {Schmiedmayer}},\ }\href {\doibase 10.2307/26462376}
  {\bibfield  {journal} {\bibinfo  {journal} {Proceedings of the National
  Academy of Sciences of the United States of America}\ }\textbf {\bibinfo
  {volume} {112}},\ \bibinfo {pages} {3866} (\bibinfo {year}
  {2015})}\BibitemShut {NoStop}%
\bibitem [{\citenamefont {Bai}\ \emph {et~al.}(2015)\citenamefont {Bai},
  \citenamefont {Harder}, \citenamefont {Chen}, \citenamefont {Fan},
  \citenamefont {Xiao},\ and\ \citenamefont {Hu}}]{bai_2015}%
  \BibitemOpen
  \bibfield  {author} {\bibinfo {author} {\bibfnamefont {L.}~\bibnamefont
  {Bai}}, \bibinfo {author} {\bibfnamefont {M.}~\bibnamefont {Harder}},
  \bibinfo {author} {\bibfnamefont {Y.~P.}\ \bibnamefont {Chen}}, \bibinfo
  {author} {\bibfnamefont {X.}~\bibnamefont {Fan}}, \bibinfo {author}
  {\bibfnamefont {J.~Q.}\ \bibnamefont {Xiao}}, \ and\ \bibinfo {author}
  {\bibfnamefont {C.-M.}\ \bibnamefont {Hu}},\ }\href {\doibase
  10.1103/PhysRevLett.114.227201} {\bibfield  {journal} {\bibinfo  {journal}
  {Phys. Rev. Lett.}\ }\textbf {\bibinfo {volume} {114}},\ \bibinfo {pages}
  {227201} (\bibinfo {year} {2015})}\BibitemShut {NoStop}%
\bibitem [{\citenamefont {Huebl}\ \emph {et~al.}(2013)\citenamefont {Huebl},
  \citenamefont {Zollitsch}, \citenamefont {Lotze}, \citenamefont {Hocke},
  \citenamefont {Greifenstein}, \citenamefont {Marx}, \citenamefont {Gross},\
  and\ \citenamefont {Goennenwein}}]{huebl_2013}%
  \BibitemOpen
  \bibfield  {author} {\bibinfo {author} {\bibfnamefont {H.}~\bibnamefont
  {Huebl}}, \bibinfo {author} {\bibfnamefont {C.~W.}\ \bibnamefont
  {Zollitsch}}, \bibinfo {author} {\bibfnamefont {J.}~\bibnamefont {Lotze}},
  \bibinfo {author} {\bibfnamefont {F.}~\bibnamefont {Hocke}}, \bibinfo
  {author} {\bibfnamefont {M.}~\bibnamefont {Greifenstein}}, \bibinfo {author}
  {\bibfnamefont {A.}~\bibnamefont {Marx}}, \bibinfo {author} {\bibfnamefont
  {R.}~\bibnamefont {Gross}}, \ and\ \bibinfo {author} {\bibfnamefont
  {S.~T.~B.}\ \bibnamefont {Goennenwein}},\ }\href {\doibase
  10.1103/PhysRevLett.111.127003} {\bibfield  {journal} {\bibinfo  {journal}
  {Phys. Rev. Lett.}\ }\textbf {\bibinfo {volume} {111}},\ \bibinfo {pages}
  {127003} (\bibinfo {year} {2013})}\BibitemShut {NoStop}%
\bibitem [{\citenamefont {Tabuchi}\ \emph {et~al.}(2014)\citenamefont
  {Tabuchi}, \citenamefont {Ishino}, \citenamefont {Ishikawa}, \citenamefont
  {Yamazaki}, \citenamefont {Usami},\ and\ \citenamefont
  {Nakamura}}]{tabuchi_2014}%
  \BibitemOpen
  \bibfield  {author} {\bibinfo {author} {\bibfnamefont {Y.}~\bibnamefont
  {Tabuchi}}, \bibinfo {author} {\bibfnamefont {S.}~\bibnamefont {Ishino}},
  \bibinfo {author} {\bibfnamefont {T.}~\bibnamefont {Ishikawa}}, \bibinfo
  {author} {\bibfnamefont {R.}~\bibnamefont {Yamazaki}}, \bibinfo {author}
  {\bibfnamefont {K.}~\bibnamefont {Usami}}, \ and\ \bibinfo {author}
  {\bibfnamefont {Y.}~\bibnamefont {Nakamura}},\ }\href {\doibase
  10.1103/PhysRevLett.113.083603} {\bibfield  {journal} {\bibinfo  {journal}
  {Phys. Rev. Lett.}\ }\textbf {\bibinfo {volume} {113}},\ \bibinfo {pages}
  {083603} (\bibinfo {year} {2014})}\BibitemShut {NoStop}%
\bibitem [{\citenamefont {Zhang}\ \emph {et~al.}(2014)\citenamefont {Zhang},
  \citenamefont {Zou}, \citenamefont {Jiang},\ and\ \citenamefont
  {Tang}}]{zhang_2014}%
  \BibitemOpen
  \bibfield  {author} {\bibinfo {author} {\bibfnamefont {X.}~\bibnamefont
  {Zhang}}, \bibinfo {author} {\bibfnamefont {C.-L.}\ \bibnamefont {Zou}},
  \bibinfo {author} {\bibfnamefont {L.}~\bibnamefont {Jiang}}, \ and\ \bibinfo
  {author} {\bibfnamefont {H.~X.}\ \bibnamefont {Tang}},\ }\href {\doibase
  10.1103/PhysRevLett.113.156401} {\bibfield  {journal} {\bibinfo  {journal}
  {Phys. Rev. Lett.}\ }\textbf {\bibinfo {volume} {113}},\ \bibinfo {pages}
  {156401} (\bibinfo {year} {2014})}\BibitemShut {NoStop}%
\bibitem [{\citenamefont {Goryachev}\ \emph {et~al.}(2014)\citenamefont
  {Goryachev}, \citenamefont {Farr}, \citenamefont {Creedon}, \citenamefont
  {Fan}, \citenamefont {Kostylev},\ and\ \citenamefont
  {Tobar}}]{goryachev_2014}%
  \BibitemOpen
  \bibfield  {author} {\bibinfo {author} {\bibfnamefont {M.}~\bibnamefont
  {Goryachev}}, \bibinfo {author} {\bibfnamefont {W.~G.}\ \bibnamefont {Farr}},
  \bibinfo {author} {\bibfnamefont {D.~L.}\ \bibnamefont {Creedon}}, \bibinfo
  {author} {\bibfnamefont {Y.}~\bibnamefont {Fan}}, \bibinfo {author}
  {\bibfnamefont {M.}~\bibnamefont {Kostylev}}, \ and\ \bibinfo {author}
  {\bibfnamefont {M.~E.}\ \bibnamefont {Tobar}},\ }\href {\doibase
  10.1103/PhysRevApplied.2.054002} {\bibfield  {journal} {\bibinfo  {journal}
  {Phys. Rev. Applied}\ }\textbf {\bibinfo {volume} {2}},\ \bibinfo {pages}
  {054002} (\bibinfo {year} {2014})}\BibitemShut {NoStop}%
\bibitem [{\citenamefont {Grigoryan}\ \emph {et~al.}(2018)\citenamefont
  {Grigoryan}, \citenamefont {Shen},\ and\ \citenamefont
  {Xia}}]{grigoryan_2018}%
  \BibitemOpen
  \bibfield  {author} {\bibinfo {author} {\bibfnamefont {V.~L.}\ \bibnamefont
  {Grigoryan}}, \bibinfo {author} {\bibfnamefont {K.}~\bibnamefont {Shen}}, \
  and\ \bibinfo {author} {\bibfnamefont {K.}~\bibnamefont {Xia}},\ }\href
  {\doibase 10.1103/PhysRevB.98.024406} {\bibfield  {journal} {\bibinfo
  {journal} {Phys. Rev. B}\ }\textbf {\bibinfo {volume} {98}},\ \bibinfo
  {pages} {024406} (\bibinfo {year} {2018})}\BibitemShut {NoStop}%
\bibitem [{\citenamefont {Harder}\ \emph {et~al.}(2018)\citenamefont {Harder},
  \citenamefont {Yang}, \citenamefont {Yao}, \citenamefont {Yu}, \citenamefont
  {Rao}, \citenamefont {Gui}, \citenamefont {Stamps},\ and\ \citenamefont
  {Hu}}]{harder_2018}%
  \BibitemOpen
  \bibfield  {author} {\bibinfo {author} {\bibfnamefont {M.}~\bibnamefont
  {Harder}}, \bibinfo {author} {\bibfnamefont {Y.}~\bibnamefont {Yang}},
  \bibinfo {author} {\bibfnamefont {B.~M.}\ \bibnamefont {Yao}}, \bibinfo
  {author} {\bibfnamefont {C.~H.}\ \bibnamefont {Yu}}, \bibinfo {author}
  {\bibfnamefont {J.~W.}\ \bibnamefont {Rao}}, \bibinfo {author} {\bibfnamefont
  {Y.~S.}\ \bibnamefont {Gui}}, \bibinfo {author} {\bibfnamefont {R.~L.}\
  \bibnamefont {Stamps}}, \ and\ \bibinfo {author} {\bibfnamefont {C.-M.}\
  \bibnamefont {Hu}},\ }\href {\doibase 10.1103/PhysRevLett.121.137203}
  {\bibfield  {journal} {\bibinfo  {journal} {Phys. Rev. Lett.}\ }\textbf
  {\bibinfo {volume} {121}},\ \bibinfo {pages} {137203} (\bibinfo {year}
  {2018})}\BibitemShut {NoStop}%
\bibitem [{\citenamefont {Grigoryan}\ and\ \citenamefont
  {Xia}(2019{\natexlab{a}})}]{grigoryan_2019}%
  \BibitemOpen
  \bibfield  {author} {\bibinfo {author} {\bibfnamefont {V.~L.}\ \bibnamefont
  {Grigoryan}}\ and\ \bibinfo {author} {\bibfnamefont {K.}~\bibnamefont
  {Xia}},\ }\href {\doibase 10.1103/PhysRevB.99.224408} {\bibfield  {journal}
  {\bibinfo  {journal} {Phys. Rev. B}\ }\textbf {\bibinfo {volume} {99}},\
  \bibinfo {pages} {224408} (\bibinfo {year} {2019}{\natexlab{a}})}\BibitemShut
  {NoStop}%
\bibitem [{\citenamefont {Bhoi}\ \emph {et~al.}(2019)\citenamefont {Bhoi},
  \citenamefont {Kim}, \citenamefont {Jang}, \citenamefont {Kim}, \citenamefont
  {Yang}, \citenamefont {Cho},\ and\ \citenamefont {Kim}}]{bhoi_2019}%
  \BibitemOpen
  \bibfield  {author} {\bibinfo {author} {\bibfnamefont {B.}~\bibnamefont
  {Bhoi}}, \bibinfo {author} {\bibfnamefont {B.}~\bibnamefont {Kim}}, \bibinfo
  {author} {\bibfnamefont {S.-H.}\ \bibnamefont {Jang}}, \bibinfo {author}
  {\bibfnamefont {J.}~\bibnamefont {Kim}}, \bibinfo {author} {\bibfnamefont
  {J.}~\bibnamefont {Yang}}, \bibinfo {author} {\bibfnamefont {Y.-J.}\
  \bibnamefont {Cho}}, \ and\ \bibinfo {author} {\bibfnamefont {S.-K.}\
  \bibnamefont {Kim}},\ }\href {\doibase 10.1103/PhysRevB.99.134426} {\bibfield
   {journal} {\bibinfo  {journal} {Phys. Rev. B}\ }\textbf {\bibinfo {volume}
  {99}},\ \bibinfo {pages} {134426} (\bibinfo {year} {2019})}\BibitemShut
  {NoStop}%
\bibitem [{\citenamefont {{Boventer}}\ \emph {et~al.}(2019)\citenamefont
  {{Boventer}}, \citenamefont {{D{\"o}rflinger}}, \citenamefont {{Wolz}},
  \citenamefont {{Mac{\^e}do}}, \citenamefont {{Lebrun}}, \citenamefont
  {{Kl{\"a}ui}},\ and\ \citenamefont {{Weides}}}]{boventer_2019}%
  \BibitemOpen
  \bibfield  {author} {\bibinfo {author} {\bibfnamefont {I.}~\bibnamefont
  {{Boventer}}}, \bibinfo {author} {\bibfnamefont {C.}~\bibnamefont
  {{D{\"o}rflinger}}}, \bibinfo {author} {\bibfnamefont {T.}~\bibnamefont
  {{Wolz}}}, \bibinfo {author} {\bibfnamefont {R.}~\bibnamefont
  {{Mac{\^e}do}}}, \bibinfo {author} {\bibfnamefont {R.}~\bibnamefont
  {{Lebrun}}}, \bibinfo {author} {\bibfnamefont {M.}~\bibnamefont
  {{Kl{\"a}ui}}}, \ and\ \bibinfo {author} {\bibfnamefont {M.}~\bibnamefont
  {{Weides}}},\ }\href@noop {} {\bibfield  {journal} {\bibinfo  {journal}
  {arXiv e-prints}\ ,\ \bibinfo {eid} {arXiv:1904.00393}} (\bibinfo {year}
  {2019})},\ \Eprint {http://arxiv.org/abs/1904.00393} {arXiv:1904.00393
  [cond-mat.mes-hall]} \BibitemShut {NoStop}%
\bibitem [{\citenamefont {{Yu}}\ \emph {et~al.}(2019)\citenamefont {{Yu}},
  \citenamefont {{Wang}}, \citenamefont {{Yuan}},\ and\ \citenamefont
  {{Xiao}}}]{yu_2019}%
  \BibitemOpen
  \bibfield  {author} {\bibinfo {author} {\bibfnamefont {W.}~\bibnamefont
  {{Yu}}}, \bibinfo {author} {\bibfnamefont {J.}~\bibnamefont {{Wang}}},
  \bibinfo {author} {\bibfnamefont {H.~Y.}\ \bibnamefont {{Yuan}}}, \ and\
  \bibinfo {author} {\bibfnamefont {J.}~\bibnamefont {{Xiao}}},\ }\href@noop {}
  {\bibfield  {journal} {\bibinfo  {journal} {arXiv e-prints}\ ,\ \bibinfo
  {eid} {arXiv:1907.06222}} (\bibinfo {year} {2019})},\ \Eprint
  {http://arxiv.org/abs/1907.06222} {arXiv:1907.06222 [cond-mat.mes-hall]}
  \BibitemShut {NoStop}%
\bibitem [{\citenamefont {Grigoryan}\ and\ \citenamefont
  {Xia}(2019{\natexlab{b}})}]{grigoryan_2019_1}%
  \BibitemOpen
  \bibfield  {author} {\bibinfo {author} {\bibfnamefont {V.~L.}\ \bibnamefont
  {Grigoryan}}\ and\ \bibinfo {author} {\bibfnamefont {K.}~\bibnamefont
  {Xia}},\ }\href {\doibase 10.1103/PhysRevB.100.014415} {\bibfield  {journal}
  {\bibinfo  {journal} {Phys. Rev. B}\ }\textbf {\bibinfo {volume} {100}},\
  \bibinfo {pages} {014415} (\bibinfo {year} {2019}{\natexlab{b}})}\BibitemShut
  {NoStop}%
\bibitem [{\citenamefont {Xu}\ \emph {et~al.}(2019)\citenamefont {Xu},
  \citenamefont {Rao}, \citenamefont {Gui}, \citenamefont {Jin},\ and\
  \citenamefont {Hu}}]{xu_2019}%
  \BibitemOpen
  \bibfield  {author} {\bibinfo {author} {\bibfnamefont {P.-C.}\ \bibnamefont
  {Xu}}, \bibinfo {author} {\bibfnamefont {J.~W.}\ \bibnamefont {Rao}},
  \bibinfo {author} {\bibfnamefont {Y.~S.}\ \bibnamefont {Gui}}, \bibinfo
  {author} {\bibfnamefont {X.}~\bibnamefont {Jin}}, \ and\ \bibinfo {author}
  {\bibfnamefont {C.-M.}\ \bibnamefont {Hu}},\ }\href {\doibase
  10.1103/PhysRevB.100.094415} {\bibfield  {journal} {\bibinfo  {journal}
  {Phys. Rev. B}\ }\textbf {\bibinfo {volume} {100}},\ \bibinfo {pages}
  {094415} (\bibinfo {year} {2019})}\BibitemShut {NoStop}%
\bibitem [{\citenamefont {{Yuan}}\ \emph {et~al.}(2019)\citenamefont {{Yuan}},
  \citenamefont {{Yan}}, \citenamefont {{Zheng}}, \citenamefont {{He}},
  \citenamefont {{Xia}},\ and\ \citenamefont {{Yung}}}]{yuan_2019}%
  \BibitemOpen
  \bibfield  {author} {\bibinfo {author} {\bibfnamefont {H.~Y.}\ \bibnamefont
  {{Yuan}}}, \bibinfo {author} {\bibfnamefont {P.}~\bibnamefont {{Yan}}},
  \bibinfo {author} {\bibfnamefont {S.}~\bibnamefont {{Zheng}}}, \bibinfo
  {author} {\bibfnamefont {Q.~Y.}\ \bibnamefont {{He}}}, \bibinfo {author}
  {\bibfnamefont {K.}~\bibnamefont {{Xia}}}, \ and\ \bibinfo {author}
  {\bibfnamefont {M.-H.}\ \bibnamefont {{Yung}}},\ }\href@noop {} {\bibfield
  {journal} {\bibinfo  {journal} {arXiv e-prints}\ ,\ \bibinfo {eid}
  {arXiv:1905.11117}} (\bibinfo {year} {2019})},\ \Eprint
  {http://arxiv.org/abs/1905.11117} {arXiv:1905.11117 [cond-mat.mes-hall]}
  \BibitemShut {NoStop}%
\bibitem [{\citenamefont {Li}\ \emph {et~al.}(2019)\citenamefont {Li},
  \citenamefont {Polakovic}, \citenamefont {Wang}, \citenamefont {Xu},
  \citenamefont {Lendinez}, \citenamefont {Zhang}, \citenamefont {Ding},
  \citenamefont {Khaire}, \citenamefont {Saglam}, \citenamefont {Divan},
  \citenamefont {Pearson}, \citenamefont {Kwok}, \citenamefont {Xiao},
  \citenamefont {Novosad}, \citenamefont {Hoffmann},\ and\ \citenamefont
  {Zhang}}]{li_2019}%
  \BibitemOpen
  \bibfield  {author} {\bibinfo {author} {\bibfnamefont {Y.}~\bibnamefont
  {Li}}, \bibinfo {author} {\bibfnamefont {T.}~\bibnamefont {Polakovic}},
  \bibinfo {author} {\bibfnamefont {Y.-L.}\ \bibnamefont {Wang}}, \bibinfo
  {author} {\bibfnamefont {J.}~\bibnamefont {Xu}}, \bibinfo {author}
  {\bibfnamefont {S.}~\bibnamefont {Lendinez}}, \bibinfo {author}
  {\bibfnamefont {Z.}~\bibnamefont {Zhang}}, \bibinfo {author} {\bibfnamefont
  {J.}~\bibnamefont {Ding}}, \bibinfo {author} {\bibfnamefont {T.}~\bibnamefont
  {Khaire}}, \bibinfo {author} {\bibfnamefont {H.}~\bibnamefont {Saglam}},
  \bibinfo {author} {\bibfnamefont {R.}~\bibnamefont {Divan}}, \bibinfo
  {author} {\bibfnamefont {J.}~\bibnamefont {Pearson}}, \bibinfo {author}
  {\bibfnamefont {W.-K.}\ \bibnamefont {Kwok}}, \bibinfo {author}
  {\bibfnamefont {Z.}~\bibnamefont {Xiao}}, \bibinfo {author} {\bibfnamefont
  {V.}~\bibnamefont {Novosad}}, \bibinfo {author} {\bibfnamefont
  {A.}~\bibnamefont {Hoffmann}}, \ and\ \bibinfo {author} {\bibfnamefont
  {W.}~\bibnamefont {Zhang}},\ }\href {\doibase 10.1103/PhysRevLett.123.107701}
  {\bibfield  {journal} {\bibinfo  {journal} {Phys. Rev. Lett.}\ }\textbf
  {\bibinfo {volume} {123}},\ \bibinfo {pages} {107701} (\bibinfo {year}
  {2019})}\BibitemShut {NoStop}%
\bibitem [{\citenamefont {Hou}\ and\ \citenamefont {Liu}(2019)}]{hou_2019}%
  \BibitemOpen
  \bibfield  {author} {\bibinfo {author} {\bibfnamefont {J.~T.}\ \bibnamefont
  {Hou}}\ and\ \bibinfo {author} {\bibfnamefont {L.}~\bibnamefont {Liu}},\
  }\href {\doibase 10.1103/PhysRevLett.123.107702} {\bibfield  {journal}
  {\bibinfo  {journal} {Phys. Rev. Lett.}\ }\textbf {\bibinfo {volume} {123}},\
  \bibinfo {pages} {107702} (\bibinfo {year} {2019})}\BibitemShut {NoStop}%
\bibitem [{\citenamefont {Troiani}(2019)}]{troiani_2019}%
  \BibitemOpen
  \bibfield  {author} {\bibinfo {author} {\bibfnamefont {F.}~\bibnamefont
  {Troiani}},\ }\href {\doibase https://doi.org/10.1016/j.physleta.2019.02.016}
  {\bibfield  {journal} {\bibinfo  {journal} {Physics Letters A}\ }\textbf
  {\bibinfo {volume} {383}},\ \bibinfo {pages} {1536 } (\bibinfo {year}
  {2019})}\BibitemShut {NoStop}%
\bibitem [{\citenamefont {Quijandr\'{\i}a}\ \emph {et~al.}(2018)\citenamefont
  {Quijandr\'{\i}a}, \citenamefont {Naether}, \citenamefont {\"Ozdemir},
  \citenamefont {Nori},\ and\ \citenamefont {Zueco}}]{quijandria_2018}%
  \BibitemOpen
  \bibfield  {author} {\bibinfo {author} {\bibfnamefont {F.}~\bibnamefont
  {Quijandr\'{\i}a}}, \bibinfo {author} {\bibfnamefont {U.}~\bibnamefont
  {Naether}}, \bibinfo {author} {\bibfnamefont {S.~K.}\ \bibnamefont
  {\"Ozdemir}}, \bibinfo {author} {\bibfnamefont {F.}~\bibnamefont {Nori}}, \
  and\ \bibinfo {author} {\bibfnamefont {D.}~\bibnamefont {Zueco}},\ }\href
  {\doibase 10.1103/PhysRevA.97.053846} {\bibfield  {journal} {\bibinfo
  {journal} {Phys. Rev. A}\ }\textbf {\bibinfo {volume} {97}},\ \bibinfo
  {pages} {053846} (\bibinfo {year} {2018})}\BibitemShut {NoStop}%
\bibitem [{\citenamefont {Zhang}\ \emph {et~al.}(2019)\citenamefont {Zhang},
  \citenamefont {Wang},\ and\ \citenamefont {You}}]{zhang_2019}%
  \BibitemOpen
  \bibfield  {author} {\bibinfo {author} {\bibfnamefont {G.-Q.}\ \bibnamefont
  {Zhang}}, \bibinfo {author} {\bibfnamefont {Y.-P.}\ \bibnamefont {Wang}}, \
  and\ \bibinfo {author} {\bibfnamefont {J.~Q.}\ \bibnamefont {You}},\ }\href
  {\doibase 10.1103/PhysRevA.99.052341} {\bibfield  {journal} {\bibinfo
  {journal} {Phys. Rev. A}\ }\textbf {\bibinfo {volume} {99}},\ \bibinfo
  {pages} {052341} (\bibinfo {year} {2019})}\BibitemShut {NoStop}%
\bibitem [{\citenamefont {Heiss}(2012)}]{heiss_2012}%
  \BibitemOpen
  \bibfield  {author} {\bibinfo {author} {\bibfnamefont {W.~D.}\ \bibnamefont
  {Heiss}},\ }\href@noop {} {\bibfield  {journal} {\bibinfo  {journal} {Journal
  of Physics A: Mathematical and Theoretical}\ }\textbf {\bibinfo {volume}
  {45}},\ \bibinfo {pages} {444016} (\bibinfo {year} {2012})}\BibitemShut
  {NoStop}%
\bibitem [{\citenamefont {Feng}\ \emph {et~al.}(2013)\citenamefont {Feng},
  \citenamefont {Xu}, \citenamefont {Fegadolli}, \citenamefont {Lu},
  \citenamefont {Oliveira}, \citenamefont {Almeida}, \citenamefont {Chen},\
  and\ \citenamefont {Scherer}}]{feng_2013}%
  \BibitemOpen
  \bibfield  {author} {\bibinfo {author} {\bibfnamefont {L.}~\bibnamefont
  {Feng}}, \bibinfo {author} {\bibfnamefont {Y.-L.}\ \bibnamefont {Xu}},
  \bibinfo {author} {\bibfnamefont {W.~S.}\ \bibnamefont {Fegadolli}}, \bibinfo
  {author} {\bibfnamefont {M.-H.}\ \bibnamefont {Lu}}, \bibinfo {author}
  {\bibfnamefont {J.~E.~B.}\ \bibnamefont {Oliveira}}, \bibinfo {author}
  {\bibfnamefont {V.~R.}\ \bibnamefont {Almeida}}, \bibinfo {author}
  {\bibfnamefont {Y.-F.}\ \bibnamefont {Chen}}, \ and\ \bibinfo {author}
  {\bibfnamefont {A.}~\bibnamefont {Scherer}},\ }\href
  {http://search.proquest.com/docview/1273814437/} {\bibfield  {journal}
  {\bibinfo  {journal} {Nature materials}\ }\textbf {\bibinfo {volume} {12}},\
  \bibinfo {pages} {108} (\bibinfo {year} {2013})}\BibitemShut {NoStop}%
\bibitem [{\citenamefont {L\"u}\ \emph {et~al.}(2015)\citenamefont {L\"u},
  \citenamefont {Jing}, \citenamefont {Ma},\ and\ \citenamefont
  {Wu}}]{lu_2015}%
  \BibitemOpen
  \bibfield  {author} {\bibinfo {author} {\bibfnamefont {X.-Y.}\ \bibnamefont
  {L\"u}}, \bibinfo {author} {\bibfnamefont {H.}~\bibnamefont {Jing}}, \bibinfo
  {author} {\bibfnamefont {J.-Y.}\ \bibnamefont {Ma}}, \ and\ \bibinfo {author}
  {\bibfnamefont {Y.}~\bibnamefont {Wu}},\ }\href {\doibase
  10.1103/PhysRevLett.114.253601} {\bibfield  {journal} {\bibinfo  {journal}
  {Phys. Rev. Lett.}\ }\textbf {\bibinfo {volume} {114}},\ \bibinfo {pages}
  {253601} (\bibinfo {year} {2015})}\BibitemShut {NoStop}%
\bibitem [{\citenamefont {Lin}\ \emph {et~al.}(2016)\citenamefont {Lin},
  \citenamefont {Pick}, \citenamefont {Lon\ifmmode~\check{c}\else
  \v{c}\fi{}ar},\ and\ \citenamefont {Rodriguez}}]{lin_2016}%
  \BibitemOpen
  \bibfield  {author} {\bibinfo {author} {\bibfnamefont {Z.}~\bibnamefont
  {Lin}}, \bibinfo {author} {\bibfnamefont {A.}~\bibnamefont {Pick}}, \bibinfo
  {author} {\bibfnamefont {M.}~\bibnamefont {Lon\ifmmode~\check{c}\else
  \v{c}\fi{}ar}}, \ and\ \bibinfo {author} {\bibfnamefont {A.~W.}\ \bibnamefont
  {Rodriguez}},\ }\href {\doibase 10.1103/PhysRevLett.117.107402} {\bibfield
  {journal} {\bibinfo  {journal} {Phys. Rev. Lett.}\ }\textbf {\bibinfo
  {volume} {117}},\ \bibinfo {pages} {107402} (\bibinfo {year}
  {2016})}\BibitemShut {NoStop}%
\bibitem [{\citenamefont {Liertzer}\ \emph {et~al.}(2012)\citenamefont
  {Liertzer}, \citenamefont {Ge}, \citenamefont {Cerjan}, \citenamefont
  {Stone}, \citenamefont {T\"ureci},\ and\ \citenamefont
  {Rotter}}]{liertzer_2012}%
  \BibitemOpen
  \bibfield  {author} {\bibinfo {author} {\bibfnamefont {M.}~\bibnamefont
  {Liertzer}}, \bibinfo {author} {\bibfnamefont {L.}~\bibnamefont {Ge}},
  \bibinfo {author} {\bibfnamefont {A.}~\bibnamefont {Cerjan}}, \bibinfo
  {author} {\bibfnamefont {A.~D.}\ \bibnamefont {Stone}}, \bibinfo {author}
  {\bibfnamefont {H.~E.}\ \bibnamefont {T\"ureci}}, \ and\ \bibinfo {author}
  {\bibfnamefont {S.}~\bibnamefont {Rotter}},\ }\href {\doibase
  10.1103/PhysRevLett.108.173901} {\bibfield  {journal} {\bibinfo  {journal}
  {Phys. Rev. Lett.}\ }\textbf {\bibinfo {volume} {108}},\ \bibinfo {pages}
  {173901} (\bibinfo {year} {2012})}\BibitemShut {NoStop}%
\bibitem [{\citenamefont {Feng}\ \emph {et~al.}(2014)\citenamefont {Feng},
  \citenamefont {Wong}, \citenamefont {Ma}, \citenamefont {Wang},\ and\
  \citenamefont {Zhang}}]{feng_2014}%
  \BibitemOpen
  \bibfield  {author} {\bibinfo {author} {\bibfnamefont {L.}~\bibnamefont
  {Feng}}, \bibinfo {author} {\bibfnamefont {Z.~J.}\ \bibnamefont {Wong}},
  \bibinfo {author} {\bibfnamefont {R.-M.}\ \bibnamefont {Ma}}, \bibinfo
  {author} {\bibfnamefont {Y.}~\bibnamefont {Wang}}, \ and\ \bibinfo {author}
  {\bibfnamefont {X.}~\bibnamefont {Zhang}},\ }\href
  {http://search.proquest.com/docview/1627077876/} {\bibfield  {journal}
  {\bibinfo  {journal} {Science (New York, N.Y.)}\ }\textbf {\bibinfo {volume}
  {346}},\ \bibinfo {pages} {972} (\bibinfo {year} {2014})}\BibitemShut
  {NoStop}%
\bibitem [{\citenamefont {Hodaei}\ \emph {et~al.}(2014)\citenamefont {Hodaei},
  \citenamefont {Mohammad-Ali}, \citenamefont {Heinrich}, \citenamefont
  {Christodoulides},\ and\ \citenamefont {Khajavikhan}}]{hodaei_2014}%
  \BibitemOpen
  \bibfield  {author} {\bibinfo {author} {\bibfnamefont {H.}~\bibnamefont
  {Hodaei}}, \bibinfo {author} {\bibfnamefont {M.}~\bibnamefont
  {Mohammad-Ali}}, \bibinfo {author} {\bibfnamefont {M.}~\bibnamefont
  {Heinrich}}, \bibinfo {author} {\bibfnamefont {D.}~\bibnamefont
  {Christodoulides}}, \ and\ \bibinfo {author} {\bibfnamefont {M.}~\bibnamefont
  {Khajavikhan}},\ }\href {http://search.proquest.com/docview/2084201860/}
  {\bibfield  {journal} {\bibinfo  {journal} {arXiv.org}\ } (\bibinfo {year}
  {2014})}\BibitemShut {NoStop}%
\bibitem [{\citenamefont {Wiersig}(2014)}]{wiersig_2014}%
  \BibitemOpen
  \bibfield  {author} {\bibinfo {author} {\bibfnamefont {J.}~\bibnamefont
  {Wiersig}},\ }\href {\doibase 10.1103/PhysRevLett.112.203901} {\bibfield
  {journal} {\bibinfo  {journal} {Phys. Rev. Lett.}\ }\textbf {\bibinfo
  {volume} {112}},\ \bibinfo {pages} {203901} (\bibinfo {year}
  {2014})}\BibitemShut {NoStop}%
\bibitem [{\citenamefont {Schreier}\ \emph {et~al.}(2015)\citenamefont
  {Schreier}, \citenamefont {Chiba}, \citenamefont {Niedermayr}, \citenamefont
  {Lotze}, \citenamefont {Huebl}, \citenamefont {Gepr\"ags}, \citenamefont
  {Takahashi}, \citenamefont {Bauer}, \citenamefont {Gross},\ and\
  \citenamefont {Goennenwein}}]{schreier_2015}%
  \BibitemOpen
  \bibfield  {author} {\bibinfo {author} {\bibfnamefont {M.}~\bibnamefont
  {Schreier}}, \bibinfo {author} {\bibfnamefont {T.}~\bibnamefont {Chiba}},
  \bibinfo {author} {\bibfnamefont {A.}~\bibnamefont {Niedermayr}}, \bibinfo
  {author} {\bibfnamefont {J.}~\bibnamefont {Lotze}}, \bibinfo {author}
  {\bibfnamefont {H.}~\bibnamefont {Huebl}}, \bibinfo {author} {\bibfnamefont
  {S.}~\bibnamefont {Gepr\"ags}}, \bibinfo {author} {\bibfnamefont
  {S.}~\bibnamefont {Takahashi}}, \bibinfo {author} {\bibfnamefont {G.~E.~W.}\
  \bibnamefont {Bauer}}, \bibinfo {author} {\bibfnamefont {R.}~\bibnamefont
  {Gross}}, \ and\ \bibinfo {author} {\bibfnamefont {S.~T.~B.}\ \bibnamefont
  {Goennenwein}},\ }\href {\doibase 10.1103/PhysRevB.92.144411} {\bibfield
  {journal} {\bibinfo  {journal} {Phys. Rev. B}\ }\textbf {\bibinfo {volume}
  {92}},\ \bibinfo {pages} {144411} (\bibinfo {year} {2015})}\BibitemShut
  {NoStop}%
\bibitem [{\citenamefont {Tserkovnyak}\ \emph {et~al.}(2005)\citenamefont
  {Tserkovnyak}, \citenamefont {Brataas}, \citenamefont {Bauer},\ and\
  \citenamefont {Halperin}}]{tserkovnyak_2005}%
  \BibitemOpen
  \bibfield  {author} {\bibinfo {author} {\bibfnamefont {Y.}~\bibnamefont
  {Tserkovnyak}}, \bibinfo {author} {\bibfnamefont {A.}~\bibnamefont
  {Brataas}}, \bibinfo {author} {\bibfnamefont {G.~E.~W.}\ \bibnamefont
  {Bauer}}, \ and\ \bibinfo {author} {\bibfnamefont {B.~I.}\ \bibnamefont
  {Halperin}},\ }\href {\doibase 10.1103/RevModPhys.77.1375} {\bibfield
  {journal} {\bibinfo  {journal} {Rev. Mod. Phys.}\ }\textbf {\bibinfo {volume}
  {77}},\ \bibinfo {pages} {1375} (\bibinfo {year} {2005})}\BibitemShut
  {NoStop}%
\bibitem [{\citenamefont {Uchida}\ \emph {et~al.}(2010)\citenamefont {Uchida},
  \citenamefont {Adachi}, \citenamefont {Ota}, \citenamefont {Nakayama},
  \citenamefont {Maekawa},\ and\ \citenamefont {Saitoh}}]{uchida_2010}%
  \BibitemOpen
  \bibfield  {author} {\bibinfo {author} {\bibfnamefont {K.-I.}\ \bibnamefont
  {Uchida}}, \bibinfo {author} {\bibfnamefont {H.}~\bibnamefont {Adachi}},
  \bibinfo {author} {\bibfnamefont {T.}~\bibnamefont {Ota}}, \bibinfo {author}
  {\bibfnamefont {H.}~\bibnamefont {Nakayama}}, \bibinfo {author}
  {\bibfnamefont {S.}~\bibnamefont {Maekawa}}, \ and\ \bibinfo {author}
  {\bibfnamefont {E.}~\bibnamefont {Saitoh}},\ }\href@noop {} {\bibfield
  {journal} {\bibinfo  {journal} {Applied Physics Letters}\ }\textbf {\bibinfo
  {volume} {97}} (\bibinfo {year} {2010})}\BibitemShut {NoStop}%
\bibitem [{\citenamefont {Holanda}\ \emph {et~al.}(2017)\citenamefont
  {Holanda}, \citenamefont {Alves~Santos}, \citenamefont
  {Rodr\'{\i}guez-Su\'arez}, \citenamefont {Azevedo},\ and\ \citenamefont
  {Rezende}}]{holanda_2017}%
  \BibitemOpen
  \bibfield  {author} {\bibinfo {author} {\bibfnamefont {J.}~\bibnamefont
  {Holanda}}, \bibinfo {author} {\bibfnamefont {O.}~\bibnamefont
  {Alves~Santos}}, \bibinfo {author} {\bibfnamefont {R.~L.}\ \bibnamefont
  {Rodr\'{\i}guez-Su\'arez}}, \bibinfo {author} {\bibfnamefont
  {A.}~\bibnamefont {Azevedo}}, \ and\ \bibinfo {author} {\bibfnamefont
  {S.~M.}\ \bibnamefont {Rezende}},\ }\href {\doibase
  10.1103/PhysRevB.95.134432} {\bibfield  {journal} {\bibinfo  {journal} {Phys.
  Rev. B}\ }\textbf {\bibinfo {volume} {95}},\ \bibinfo {pages} {134432}
  (\bibinfo {year} {2017})}\BibitemShut {NoStop}%
\bibitem [{\citenamefont {Safranski}\ \emph {et~al.}(2017)\citenamefont
  {Safranski}, \citenamefont {Barsukov}, \citenamefont {Lee}, \citenamefont
  {Schneider}, \citenamefont {Jara}, \citenamefont {Smith}, \citenamefont
  {Chang}, \citenamefont {Lenz}, \citenamefont {Lindner}, \citenamefont
  {Tserkovnyak},\ and\ \citenamefont {Krivorotov}}]{safranski_2017}%
  \BibitemOpen
  \bibfield  {author} {\bibinfo {author} {\bibfnamefont {C.}~\bibnamefont
  {Safranski}}, \bibinfo {author} {\bibfnamefont {I.}~\bibnamefont {Barsukov}},
  \bibinfo {author} {\bibfnamefont {H.}~\bibnamefont {Lee}}, \bibinfo {author}
  {\bibfnamefont {T.}~\bibnamefont {Schneider}}, \bibinfo {author}
  {\bibfnamefont {A.}~\bibnamefont {Jara}}, \bibinfo {author} {\bibfnamefont
  {A.}~\bibnamefont {Smith}}, \bibinfo {author} {\bibfnamefont
  {H.}~\bibnamefont {Chang}}, \bibinfo {author} {\bibfnamefont
  {K.}~\bibnamefont {Lenz}}, \bibinfo {author} {\bibfnamefont {J.}~\bibnamefont
  {Lindner}}, \bibinfo {author} {\bibfnamefont {Y.}~\bibnamefont
  {Tserkovnyak}}, \ and\ \bibinfo {author} {\bibfnamefont {I.}~\bibnamefont
  {Krivorotov}},\ }\href {http://search.proquest.com/docview/1923278830/}
  {\bibfield  {journal} {\bibinfo  {journal} {Nature Communications}\ }\textbf
  {\bibinfo {volume} {8}},\ \bibinfo {pages} {1} (\bibinfo {year}
  {2017})}\BibitemShut {NoStop}%
\bibitem [{\citenamefont {Chen}\ \emph {et~al.}(2016)\citenamefont {Chen},
  \citenamefont {Dumas}, \citenamefont {Eklund}, \citenamefont {Muduli},
  \citenamefont {Houshang}, \citenamefont {Awad}, \citenamefont {Durrenfeld},
  \citenamefont {Malm}, \citenamefont {Rusu},\ and\ \citenamefont
  {Akerman}}]{chen_2016}%
  \BibitemOpen
  \bibfield  {author} {\bibinfo {author} {\bibfnamefont {T.}~\bibnamefont
  {Chen}}, \bibinfo {author} {\bibfnamefont {R.~K.}\ \bibnamefont {Dumas}},
  \bibinfo {author} {\bibfnamefont {A.}~\bibnamefont {Eklund}}, \bibinfo
  {author} {\bibfnamefont {P.~K.}\ \bibnamefont {Muduli}}, \bibinfo {author}
  {\bibfnamefont {A.}~\bibnamefont {Houshang}}, \bibinfo {author}
  {\bibfnamefont {A.~A.}\ \bibnamefont {Awad}}, \bibinfo {author}
  {\bibfnamefont {P.}~\bibnamefont {Durrenfeld}}, \bibinfo {author}
  {\bibfnamefont {B.~G.}\ \bibnamefont {Malm}}, \bibinfo {author}
  {\bibfnamefont {A.}~\bibnamefont {Rusu}}, \ and\ \bibinfo {author}
  {\bibfnamefont {J.}~\bibnamefont {Akerman}},\ }\href {\doibase
  10.1109/JPROC.2016.2554518} {\bibfield  {journal} {\bibinfo  {journal}
  {Proceedings of the IEEE}\ }\textbf {\bibinfo {volume} {104}},\ \bibinfo
  {pages} {1919} (\bibinfo {year} {2016})}\BibitemShut {NoStop}%
\bibitem [{\citenamefont {Hamadeh}\ \emph {et~al.}(2014)\citenamefont
  {Hamadeh}, \citenamefont {d'Allivy Kelly}, \citenamefont {Hahn},
  \citenamefont {Meley}, \citenamefont {Bernard}, \citenamefont {Molpeceres},
  \citenamefont {Naletov}, \citenamefont {Viret}, \citenamefont {Anane},
  \citenamefont {Cros}, \citenamefont {Demokritov}, \citenamefont {Prieto},
  \citenamefont {Munoz}, \citenamefont {de~Loubens},\ and\ \citenamefont
  {Klein}}]{hamadeh_2014}%
  \BibitemOpen
  \bibfield  {author} {\bibinfo {author} {\bibfnamefont {A.}~\bibnamefont
  {Hamadeh}}, \bibinfo {author} {\bibfnamefont {O.}~\bibnamefont {d'Allivy
  Kelly}}, \bibinfo {author} {\bibfnamefont {C.}~\bibnamefont {Hahn}}, \bibinfo
  {author} {\bibfnamefont {H.}~\bibnamefont {Meley}}, \bibinfo {author}
  {\bibfnamefont {R.}~\bibnamefont {Bernard}}, \bibinfo {author} {\bibfnamefont
  {A.~H.}\ \bibnamefont {Molpeceres}}, \bibinfo {author} {\bibfnamefont
  {V.~V.}\ \bibnamefont {Naletov}}, \bibinfo {author} {\bibfnamefont
  {M.}~\bibnamefont {Viret}}, \bibinfo {author} {\bibfnamefont
  {A.}~\bibnamefont {Anane}}, \bibinfo {author} {\bibfnamefont
  {V.}~\bibnamefont {Cros}}, \bibinfo {author} {\bibfnamefont {S.~O.}\
  \bibnamefont {Demokritov}}, \bibinfo {author} {\bibfnamefont {J.~L.}\
  \bibnamefont {Prieto}}, \bibinfo {author} {\bibfnamefont {M.}~\bibnamefont
  {Munoz}}, \bibinfo {author} {\bibfnamefont {G.}~\bibnamefont {de~Loubens}}, \
  and\ \bibinfo {author} {\bibfnamefont {O.}~\bibnamefont {Klein}},\ }\href
  {\doibase 10.1103/PhysRevLett.113.197203} {\bibfield  {journal} {\bibinfo
  {journal} {Phys. Rev. Lett.}\ }\textbf {\bibinfo {volume} {113}},\ \bibinfo
  {pages} {197203} (\bibinfo {year} {2014})}\BibitemShut {NoStop}%
\bibitem [{\citenamefont {Sklenar}\ \emph {et~al.}(2015)\citenamefont
  {Sklenar}, \citenamefont {Zhang}, \citenamefont {Jungfleisch}, \citenamefont
  {Jiang}, \citenamefont {Chang}, \citenamefont {Pearson}, \citenamefont {Wu},
  \citenamefont {Ketterson},\ and\ \citenamefont {Hoffmann}}]{sklenar_2015}%
  \BibitemOpen
  \bibfield  {author} {\bibinfo {author} {\bibfnamefont {J.}~\bibnamefont
  {Sklenar}}, \bibinfo {author} {\bibfnamefont {W.}~\bibnamefont {Zhang}},
  \bibinfo {author} {\bibfnamefont {M.~B.}\ \bibnamefont {Jungfleisch}},
  \bibinfo {author} {\bibfnamefont {W.}~\bibnamefont {Jiang}}, \bibinfo
  {author} {\bibfnamefont {H.}~\bibnamefont {Chang}}, \bibinfo {author}
  {\bibfnamefont {J.~E.}\ \bibnamefont {Pearson}}, \bibinfo {author}
  {\bibfnamefont {M.}~\bibnamefont {Wu}}, \bibinfo {author} {\bibfnamefont
  {J.~B.}\ \bibnamefont {Ketterson}}, \ and\ \bibinfo {author} {\bibfnamefont
  {A.}~\bibnamefont {Hoffmann}},\ }\href {\doibase 10.1103/PhysRevB.92.174406}
  {\bibfield  {journal} {\bibinfo  {journal} {Phys. Rev. B}\ }\textbf {\bibinfo
  {volume} {92}},\ \bibinfo {pages} {174406} (\bibinfo {year}
  {2015})}\BibitemShut {NoStop}%
\bibitem [{\citenamefont {Walls}(1994)}]{walls_1994}%
  \BibitemOpen
  \bibfield  {author} {\bibinfo {author} {\bibfnamefont {D.~F.}\ \bibnamefont
  {Walls}},\ }\href@noop {} {\emph {\bibinfo {title} {Quantum optics}}}\
  (\bibinfo  {publisher} {Springer},\ \bibinfo {address} {Berlin ; New York},\
  \bibinfo {year} {1994})\BibitemShut {NoStop}%
\bibitem [{\citenamefont {Clerk}\ \emph {et~al.}(2010)\citenamefont {Clerk},
  \citenamefont {Devoret}, \citenamefont {Girvin}, \citenamefont {Marquardt},\
  and\ \citenamefont {Schoelkopf}}]{clerk_2010}%
  \BibitemOpen
  \bibfield  {author} {\bibinfo {author} {\bibfnamefont {A.~A.}\ \bibnamefont
  {Clerk}}, \bibinfo {author} {\bibfnamefont {M.~H.}\ \bibnamefont {Devoret}},
  \bibinfo {author} {\bibfnamefont {S.~M.}\ \bibnamefont {Girvin}}, \bibinfo
  {author} {\bibfnamefont {F.}~\bibnamefont {Marquardt}}, \ and\ \bibinfo
  {author} {\bibfnamefont {R.~J.}\ \bibnamefont {Schoelkopf}},\ }\href
  {\doibase 10.1103/RevModPhys.82.1155} {\bibfield  {journal} {\bibinfo
  {journal} {Rev. Mod. Phys.}\ }\textbf {\bibinfo {volume} {82}},\ \bibinfo
  {pages} {1155} (\bibinfo {year} {2010})}\BibitemShut {NoStop}%
\end{thebibliography}
\end{document}